\documentclass[usenatbib,usegraphicx]{mn2e}

\usepackage{graphicx}
\usepackage{url}

\title[CFHTLenS: Photometric redshifts]{CFHTLenS: Improving the
  quality of photometric redshifts with precision
  photometry\thanks{Based on observations obtained with
    MegaPrime/MEGACAM, a joint project of CFHT and CEA/DAPNIA, at the
    Canada-France-Hawaii Telescope (CFHT) which is operated by the
    National Research Council (NRC) of Canada, the Institut National
    des Sciences de l'Univers of the Centre National de la Recherche
    Scientifique (CNRS) of France, and the University of Hawaii. This
    work is based in part on data products produced at TERAPIX and the
    Canadian Astronomy Data Centre as part of the Canada-France-Hawaii
    Telescope Legacy Survey, a collaborative project of NRC and
    CNRS.}}

\author[H.~Hildebrandt et al.]{H.~Hildebrandt,$^{1,2}$\thanks{Email: hendrik@phas.ubc.ca}
T.~Erben,$^2$
K.~Kuijken,$^3$
L.~van~Waerbeke,$^1$
C.~Heymans,$^4$
\newauthor
J.~Coupon,$^5$
J.~Benjamin,$^1$
C.~Bonnett,$^6$
L.~Fu,$^7$
H.~Hoekstra,$^3$
T.~D.~Kitching,$^4$
\newauthor
Y.~Mellier,$^{8,9}$
L.~Miller,$^{10}$
M.~Velander,$^{3,10}$
M.~J.~Hudson,$^{11,12}$
B.~T.~P.~Rowe,$^{13}$
\newauthor
T.~Schrabback,$^{14,3}$
E.~Semboloni,$^3$
N.~Ben\'itez$^{15}$
\\
$^1$University of British Columbia, Department of Physics and Astronomy, 6224 Agricultural Road, Vancouver, B.C. V6T 1Z1, Canada\\
$^2$Argelander-Institut f\"ur Astronomie, Auf dem H\"ugel 71, 53121 Bonn, Germany\\
$^3$Leiden Observatory, Leiden University, Niels Bohrweg 2, 2333 CA Leiden, The Netherlands\\
$^4$Scottish Universities Physics Alliance, Institute for Astronomy, University of Edinburgh, Royal Observatory, Blackford Hill,\\ Edinburgh EH9 3HJ, UK\\
$^5$Astronomical Institute, Graduate School of Science, Tohoku University, Sendai 980-8578, Japan\\
$^6$Institut de Ci\`encies de l'Espai, CSIC/IEEC, F. de Ci\`encies, Torre C5 par-2,  Barcelona 08193, Spain\\
$^7$Key Lab for Astrophysics, Shanghai Normal University, 100 Guilin Road, 200234, Shanghai, PR China\\
$^8$Universit\'e Pierre et Marie Curie-Paris 6, Institut d'Astrophysique de Paris, 98 bis Boulevard Arago, F-75014 Paris, France\\
$^9$CNRS, UMR 7095, Institut d’Astrophysique de Paris, 98 bis Boulevard Arago, F-75014 Paris, France\\
$^{10}$Department of Physics, Oxford University, Keble Road, Oxford OX1 3RH, UK\\
$^{11}$Department of Physics and Astronomy, University of Waterloo, Waterloo, ON N2L 3G1, Canada\\
$^{12}$Perimeter Institute for Theoretical Physics, 31 Caroline St N, Waterloo, Ontario, N2L 2Y5, Canada\\
$^{13}$Department of Physics and Astronomy, University College London, Gower Street, London WC1E 6BT, UK\\
$^{14}$Kavli Institute for Particle Astrophysics and Cosmology, Stanford University, 382 Via Pueblo Mall, Stanford, CA 94305-4060, USA\\
$^{15}$Instituto de Astrof\'isica de Andaluc\'ia (CSIC), C/Camino Bajo de Hu\'etor 24, Granada 18008, Spain\\
}

\date{Released 2011 Xxxxx XX}

\pagerange{\pageref{firstpage}--\pageref{lastpage}} \pubyear{2011}

\begin{document}
\label{firstpage}

\maketitle
\clearpage
\begin{abstract}
Here we present the results of various approaches to measure accurate
colours and photometric redshifts (photo-$z$'s) from wide-field
imaging data. We use data from the Canada-France-Hawaii-Telescope
Legacy Survey (CFHTLS) which have been re-processed by the CFHT
Lensing Survey (CFHTLenS) team in order to carry out a number of weak
gravitational lensing studies. An emphasis is put on the correction of
systematic effects in the photo-$z$'s arising from the different Point
Spread Functions (PSF) in the five optical bands. Different ways of
correcting these effects are discussed and the resulting photo-$z$
accuracies are quantified by comparing the photo-$z$'s to large
spectroscopic redshift (spec-$z$) data sets. Careful homogenisation of
the PSF between bands leads to increased overall accuracy of
photo-$z$'s. The gain is particularly pronounced at fainter magnitudes
where galaxies are smaller and flux measurements are affected more by
PSF-effects. We discuss ways of defining more secure subsamples of
galaxies as well as a shape- and colour-based star-galaxy separation
method, and we present redshift distributions for different magnitude
limits. We also study possible re-calibrations of the photometric
zeropoints (ZPs) with the help of galaxies with known spec-$z$'s. We
find that if PSF-effects are properly taken into account, a
re-calibration of the ZPs becomes much less important suggesting that
previous such re-calibrations described in the literature could in
fact be mostly corrections for PSF-effects rather than corrections for
real inaccuracies in the ZPs. The implications of this finding for
future surveys like KiDS, DES, LSST, or Euclid are mixed. On the one
hand, ZP re-calibrations with spec-$z$'s might not be as accurate as
previously thought. On the other hand, careful PSF homogenisation
might provide a way out and yield accurate, homogeneous photometry
without the need for full spectroscopic coverage. This is the first
paper in a series describing the technical aspects of CFHTLenS.
\end{abstract}

\section{Introduction}
\label{sec:intro}
Estimating distances of celestial objects has always been one of the
major technical aspects in observational astronomy. Whenever
approximate redshifts of a very large number of faint extragalactic
objects are needed the estimation of redshifts (and hence distances)
from colours, also termed photometric redshifts \citep[photo-$z$'s;
  see
  e.g.][]{1962IAUS...15..390B,1982ApJ...257L..57P,1985AJ.....90..418K,1986ApJ...303..154L,1995AJ....110.2655C,1999ASPC..191....3K,2000ApJ...536..571B,2000A&A...363..476B,2001A&A...365..660W,2003AJ....125..580C,2004PASP..116..345C,2006A&A...457..841I,2008A&A...480..703H,2010A&A...523A..31H,2009A&A...500..981C},
represent the only practical solution. Over the last few decades this
technique has become increasingly important in extragalactic
studies. Cosmological observations, inherently statistical in nature,
particularly benefit from the availability of redshifts for millions
of objects over large cosmological volumes.

An example is weak gravitational lensing \citep[WL; for reviews
  see][]{2001PhR...340..291B,2008PhR...462...67M,2008ARNPS..58...99H},
which has been established as an important tool to study the dark
sector of the Universe. The first WL detection of a cluster of
galaxies was made by \cite{1990ApJ...349L...1T}. Thanks to the steady
progress in analysis tools and better knowledge of the source redshift
distribution WL studies have become an important tool to calibrate the
masses of individual galaxy clusters
\citep{2007MNRAS.379..317H,2008MNRAS.385.1431H,2010PASJ...62..811O,2011ApJ...726...48H,2011ApJ...737...59J}. By
stacking the signals of many lenses, the average properties of
clusters \citep{2007arXiv0709.1159J,2011ApJ...733L..30H} and groups
\citep{2001ApJ...548L...5H,2005ApJ...634..806P,2010ApJ...709...97L} or
even galaxies
\citep[e.g.][]{1996ApJ...466..623B,1998ApJ...503..531H,2004ApJ...606...67H,2006MNRAS.371L..60H,2006MNRAS.368..715M,2007ApJ...669...21P,2011arXiv1104.0928L,2011A&A...534A..14V}
can be studied. Furthermore, the WL effect of the large scale
structure (LSS) of the Universe, called cosmic shear
\citep{2000A&A...358...30V,2000astro.ph..3338K,2000MNRAS.318..625B,2000Natur.405..143W,2006A&A...452...51S,2006ApJ...647..116H,2007A&A...468..859H,2007A&A...468..823S,2010A&A...516A..63S,2007ApJS..172..239M,2008A&A...479....9F},
has been identified as one of the most promising probes of the effects
of dark energy \citep[DE;][]{2006astro.ph..9591A,2006ewg3.rept.....P}.

Since the lensing signals observed in these cases depend directly on
the distances of the lenses and sources it is important to have an
accurate knowledge of the lens-source geometry through knowing the
redshifts of the objects. Modern WL surveys are designed in such a way
that simultaneous measurements of the WL observables as well as
photo-$z$'s are possible. In particular large, future imaging surveys
that will cover a fair fraction of the extragalactic sky - like KiDS,
DES, LSST, and Euclid - require extremely accurate photo-$z$'s to
reach a systematic accuracy in the WL measurements that does not
compromise the survey's statistical power. Different ways of achieving
this goal have been discussed in the literature.

In the foreseeable future all of these surveys will rely on
ground-based multi-colour photometry, which means that atmospheric
effects have to be corrected for. In this paper we present advanced
techniques to arrive at homogeneous photo-$z$'s from inherently
inhomogeneous, ground-based survey data. For this we use the most
powerful WL survey to date, the Canada France Hawaii Telescope Legacy
Survey (CFHTLS). Being a ground-based survey, the CFHTLS involves some
unavoidable inhomogeneities, e.g. in terms of seeing, atmospheric
extinction, etc. The CFHT Lensing Survey
(CFHTLenS\footnote{\url{http://www.cfhtlens.org}}) team was formed to
provide a reduction and analysis of the CFHTLS data optimised for WL
science and addressing these challenges. The higher-level requirements
on the data to measure accurate shapes and redshifts of tens of
millions of galaxies made this 'lensing-quality' reduction
necessary.

In this first paper of a series we present the multi-colour photometry
and the resulting photo-$z$'s upon which the future CFHTLenS science
projects will be based. In Sect.~\ref{sec:data} the CFHTLS data set
and the CFHTLenS data reduction are presented and compared to the
public data available from the TERAPIX team. Section~\ref{sec:PSFhomo}
deals with the crucially important correction for atmospheric effects
needed for accurate multi-colour photometry. In
Sect.~\ref{sec:catalogues} our strategy to extract catalogues from the
images in five different photometric bands is described. The photo-$z$
estimation is then presented in Sect.~\ref{sec:photoz} along with the
results of the comparisons between photo-$z$'s and spectroscopic
redshifts. Conclusions are presented in Sect.~\ref{sec:conclusions}.

\section{Data set}
\label{sec:data}
\subsection{The CFHTLS-Wide}
\label{sec:CFHTLS}
The Wide component of the Canada-France-Hawaii Telescope Legacy Survey
(CFHTLS-Wide) commenced in mid-2003 and completed observations in
early 2009.  In more than 2300 hours of dark and grey time over these
5 and a half years, the CFHTLS-Wide imaged 172 one square degree
MEGACAM fields in five filters $u,g,r,i,z$ to a $5\sigma$ point source
limiting magnitude of $ i_{\rm AB}\approx25.5$.  The data span four
contiguous fields; W1($\sim 63.8$ sq. deg.), W2($\sim 22.6$ sq. deg.),
W3($\sim 44.2$ sq. deg.) and W4($\sim 23.3$ sq. deg.) totalling 154
sq. deg. once the overlap regions are accounted for.  W1, W2 and W4
are equatorial fields with W1 and W4 containing the VVDS and VIPERS
spectroscopic surveys.  W3 is a northern field containing the extended
Groth Strip DEEP2 spectroscopic survey.  A detailed report of the full
CFHTLS Deep and Wide surveys can be found in the TERAPIX CFHTLS T0006
release document\citep{T0006}.

The CFHTLS-Wide was optimised for the study of weak gravitational
lensing for which the crucial observables are the shape of resolved
galaxies as well as their redshifts. The observing strategy was
therefore to reserve the best seeing conditions with $\theta<0\farcs8$ for
the lensing $i$-band filter and follow-up with the other bands in the
poorer seeing conditions. That is also the reason why the $i$-band is
our primary object detection band (see Sect.~\ref{sec:catalogues}).

\subsection{The CFHTLenS data reduction}
\label{sec:reduction}
The data reduction was conducted with the THELI pipeline
\citep{2003A&A...407..869S,2005AN....326..432E} following the
procedures outlined in \cite{2009A&A...493.1197E}. We briefly
summarise the most important differences in data processing between
the CARS project, detailed in \cite{2009A&A...493.1197E} and
\cite{2009A&A...498..725H}, and the current CFHTLenS data set. A more
detailed description will be given in another publication of this
series.

The CFHTLenS project makes use of the complete CFHTLS-Wide data
set. This includes five-colour coverage of 172 square degrees of
high-quality data subdivided in four patches W1-W4 (see
Sect.~\ref{sec:CFHTLS}).  In addition we make use of the CFHTLS
Pre-Survey which densely covers the complete survey area with shortly
exposed $r$-band images. This Pre-Survey was acquired to optimise the
astrometric calibration for the main science data. Similarly, to
improve the photometric calibration of CFHTLS, the survey area was
(re-)observed in a sparse grid under photometric conditions during the
year 2008 (CFHT program RunIDs 08AL99 and 08BL99).

All CFHT MEGACAM images are initially processed using the Elixir
software at the Canadian Astronomical Data Centre (CADC) and it is
this archived data that we use in the CFHTLenS project. The current
work and all other CFHTLenS publications use the CFHTLS-Wide images,
the astrometric Pre-Survey data and additional photometric data that
were available at CADC on January 15, 2009. In total, the set contains
7997 Elixir processed CFHT MEGACAM images.

\begin{figure}
\includegraphics[width=\hsize]{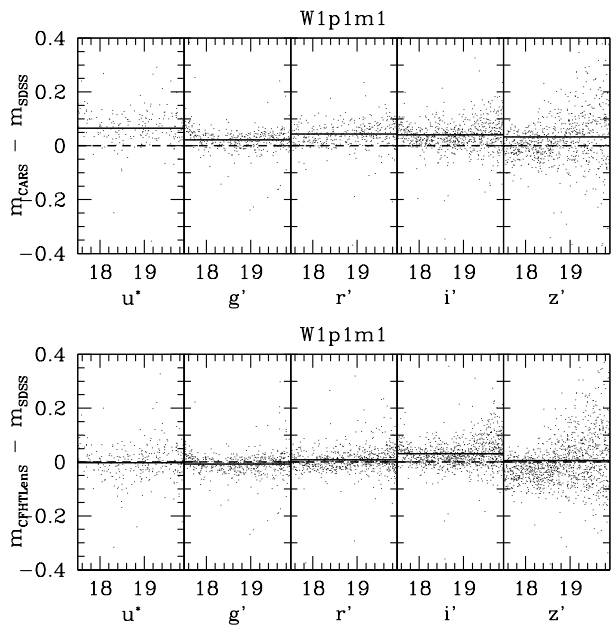}
\caption{Photometric comparison between SDSS-DR5 and CARS \citep[top;
    see][]{2009A&A...493.1197E} and SDSS-DR5 and CFHTLenS (bottom) in
  the five different bands in one field of $\approx1$~sq.~deg. The
  solid horizontal lines represent the average difference over this
  magnitude interval.}
\label{fig:photometry}
\end{figure}

While we processed the data on a per-pointing level in the CARS
project, we performed all calibrations on a per-patch level for
CFHTLenS.  The inclusion of all available data, especially the
astrometric Pre-Survey and the photometric (re-)calibration,
significantly improved the homogeneity of our data. Most important for
this work is our improvement in photometric calibration. In
\cite{2009A&A...493.1197E} we quoted the RMS uncertainty of our
relative photometric calibration between fields as $\sigma_{abs,
  g'r'i'} \approx 0.01-0.04$ mag, $\sigma_{abs, z'} \approx 0.03-0.05$
mag and $\sigma_{abs, u^*} \approx 0.15$ mag. In CFHTLenS we now reach
$\sigma \approx 0.01-0.03$ mag in all
passbands. Figure~\ref{fig:photometry} compares magnitudes in the
CFHTLenS field W1p1m1 with the magnitudes taken from the fifth data
release of the Sloan Digital Sky Survey
\citep[SDSS-DR5;][]{2007ApJS..172..634A}. For comparison the plot also
shows this comparison for the corresponding CARS field \cite[see also
  Fig.~A.7 of][]{2009A&A...493.1197E}.

\subsection{The TERAPIX T0006 catalogue}
\label{sec:T0006}
The first reduction of the CFHTLS-Wide data was performed by
TERAPIX\footnote{\url{http://terapix.iap.fr/}}. The data were
distributed in several public releases. The most recent release, T0006
\citep[November 15, 2010; ][]{T0006}, includes photo-$z$'s which were
estimated in the same way as described in \cite{2009A&A...500..981C}
for the T0004 release \citep[see also][for more details on the T0006
  photo-$z$'s]{2011arXiv1107.0616C} with the \emph{Le Phare} photo-$z$
code \citep{2002MNRAS.329..355A,2006A&A...457..841I}. However, the
photometric calibration improved significantly in the latest release
and now covers the whole CFHTLS-Wide area. We will compare the
photo-$z$'s from the T0006 catalogues to CFHTLenS in the following
sections.

The most notable differences in the multi-colour photometry compared
to CFHTLenS are:
\begin{enumerate}
\item The T0006 images are median stacks whereas the CFHTLenS images
  are mean stacks \citep{2009A&A...493.1197E}. While median stacks
  allow for easier rejection of image defects (cosmic rays,
  reflections, asteroids, etc.) mean stacks are less noisy and hence
  lead to deeper images and more precise photometry.\footnote{A lot of
    effort was invested to automatically mask image defects in the
    CFHTLenS data reduction prior to the stacking. This leads to very
    clean images which can be averaged directly superseding the more
    noisy median (or also trimmed-mean) procedures. The standard error
    of the median is $\approx 25$ per cent larger than the standard
    error of the mean, if the noise distribution is Gaussian, leading
    to a loss of $\approx 0.24$mag of depth.}
\item T0006 does not implement PSF homogenisation (a procedure
  correcting for the different seeings in the different bands as well
  as at different positions on the image described in
  Sect.~\ref{sec:PSFhomo}) before catalogue extraction.
\item  T0006 measures magnitudes in elliptical apertures
  \citep{1980ApJS...43..305K} instead of isophotal apertures, which we
  use for CFHTLenS (see Sect.~\ref{sec:catalogues}).
\item The photometric zeropoints were re-calibrated with the help
  of a large sample of spectroscopic redshifts. See
  Sect.~\ref{sec:recalib} for a detailed discussion of the benefits
  and dangers of such a re-calibration.
\end{enumerate}

\section{PSF homogenisation}
\label{sec:PSFhomo}
 For photo-$z$'s we are particularly interested in measuring accurate
 colours of objects. For point sources this is a clearly defined
 problem. For extended objects, however, there is no unique definition
 of a colour and it depends on circumstances as to which definition is
 the most useful. For our purpose, to use the colours to estimate
 photo-$z$'s (see Sect.~\ref{sec:photoz}) we need colours that best
 match those modelled from SED templates. As we explain below this
 requires identical physical apertures\footnote{Here the term
   'physical' refers to apertures which cover the same physical parts
   of an object in different bands, i.e. the same area on the sky
   before blurring by the atmosphere.} on the sky in the different
 bands. An extreme choice would be to define colour as the difference
 in total magnitude of an object in different bands: but since objects
 can have very different physical extent in different bands this
 approach could lead to vastly different colours compared to the
 matched apertures, and is not optimal in a terms of signal to noise.

When estimating photo-$z$'s we compare the observed colours of an object
to colours modelled from the convolution of SED templates with
instrument response curves. These SED templates are either based on
observations (empirical templates) or on synthetic stellar evolution
models (synthetic templates). Both approaches yield model colours for
a particular stellar population, and it is therefore important that
our observations yield colours that correspond to the same set of
stars in each band, i.e., that they represent the same physical
aperture on the galaxy.

Matching the physical apertures is complicated by the fact that
different images taken through different filter bands will invariably
show different degrees of blurring, expressed as the point-spread
function (PSF) and measurable as the observed shape of the stellar
images in the field. These PSF differences can arise from different
ambient conditions during the observations, or from chromatic effects
in the atmosphere and optics of the telescope. Regardless of their
origin they have the effect that identical apertures in different
images of the same part of the sky do not represent the same physical
part of a source. Not only different PSF sizes (seeing), but also the
more subtle effects of PSF anisotropy, especially with a
  prime-focus camera like MEGACAM, will affect the
photometry. Compensating for the PSF differences between bands can be
done in two ways: either by adapting the apertures to compensate for
the different PSFs, or by manipulating the images so as to arrive at
images with the same PSFs. Here we will follow the latter approach,
because it is independent of the object size, and leads to more
homogeneous photometry across the survey. This involves constructing a
suitable pixel convolution kernel for each image.

\subsection{Constant Gaussian convolution kernel: \emph{global}}
In this method we assume for simplicity that the PSF can be described
by a single Gaussian with width $\sigma_{\rm PSF}$. Under this
assumption one can convolve an image in band $X$ with a 2-dimensional
Gaussian filter function of width
\begin{equation}
\sigma_{{\rm filter},  X}=\sqrt{\sigma_{\rm PSF, worst}^2-\sigma_{{\rm PSF}, X}^2}\,,
\end{equation}

to arrive at an image with a PSF size which matches the PSF size of
the image with the worst seeing in a given set, $\sigma_{\rm PSF,
  worst}$. This method implies that the PSF size does not change with
position on an image (or at least that the variation with position is
the same in all bands)\footnote {However, it should be noted that
  PSF-effects can have a collective effect changing the average
  properties of galaxies as a function of position if they are not
  corrected for. We defer an analysis of these PSF-photo-$z$
  correlations to a forthcoming paper.}, an assumption that is not
necessarily true for contemporary wide-field imaging cameras like
MEGACAM on the CFHT. In the following we will call this approach
\emph{global} PSF homogenisation. It is the same approach as used in
\cite{2006A&A...452.1121H,2007A&A...462..865H,2009A&A...498..725H} and
\cite{2009A&A...493.1197E} and is computationally straightforward.

\subsection{Gaussianisation of the PSF with a spatially varying kernel: \emph{local} }
\label{sec:PSFhomo_local}
In the \emph{local} approach we drop the assumptions that the initial
PSF is Gaussian and position-independent, and construct a convolution
kernel that is designed to make the PSF Gaussian everywhere, with the
same width. Obviously this requires a non-Gaussian convolution kernel
that changes with position on the image. Here we model the PSF and
convolution kernels using the shapelet formalism
\citep{2003MNRAS.338...35R,2003MNRAS.338...48R,2006A&A...456..827K}---essentially,
each source is described as a sum of two-dimensional Gauss-Hermite
functions. This method is closely related to the GaaP photometric
package described in \cite{2008A&A...482.1053K}.

First, a catalogue is extracted from the $i$-band image with the
SExtractor software \citep{1996A&AS..117..393B}. Stars are selected in
a magnitude vs. size diagram from the $i$-band data. In each of the
five bands, the images of these stars are then modelled by sums of
shapelets up to tenth order, through least-squares fitting of the
pixel values. The size of the PSF, which is needed to scale the
shapelets, is also taken from the initial SExtractor catalogue.

Next two-dimensional, fifth-order polynomials are fitted to the
coefficients of the shapelet expansion to describe the variation of
the PSF over the MEGACAM field in each band. From these high-order
analytic descriptions of the PSF variations in each of the five bands,
the convolution kernel is then created.  The goal is to produce five
images which have exactly the same Gaussian PSF over the whole
field. To avoid deconvolution and the associated noise amplification,
the target size for the final Gaussian PSF is chosen to match the
largest PSF size found in any position in any of the five bands of a
field. Within the shapelet formalism, it is easy to calculate a
convolution kernel that transforms the modelled PSF into the Gaussian
target PSF at each position in each band, and to perform the
convolution in Fourier space. More details can be found in
appendix~\ref{appA}.

In the following we will call this approach \emph{local} PSF
Gaussianisation.

\subsection{General remarks}
\label{sec:PSFhomo_general}
It is clear that the benefits of PSF homogenisation are most
pronounced if the changes in PSF size (or also PSF shape) are large
between bands.  We will compare both the \emph{global} and the
\emph{local} approaches to the case where no PSF homogenisation is
performed. This latter case will be referred to with the label
\emph{none}. See Table~\ref{tab:PSF_schemes} for a summary of the
different schemes to homogenise the PSF. The photo-$z$'s in the T0006
catalogues (see Sect.~\ref{sec:T0006}), which were estimated from
multi-colour photometry on median stacks that had not been corrected
for PSF-effects, will be referred to by the label \emph{T0006}.

\begin{table*}
\begin{minipage}[t]{\textwidth}
\caption{Different schemes to homogenise the PSF and their properties. Note that the PSF size at a random position in the \emph{global} scheme is only approximately constant between bands because the intrinsic PSF size changes from the centre to the edge of the MEGACAM mosaic. Using a constant convolution kernel for PSF homogenisation between bands will only correct the PSF in one position (here chosen to be the centre of the image) and leave residuals at other positions. However, as we show in Sect.~\ref{sec:accuracy}, these residuals are small enough to make the photo-$z$'s based on the \emph{global} photometry superior to the ones based on the \emph{none} or \emph{T0006} photometry (but not as accurate as the ones based on the \emph{local} photometry).}
\centering
\renewcommand{\footnoterule}{}  
\begin{tabular}{lcccc}
\hline
\hline
Scheme     & PSF size constant    & PSF size constant     & PSF size \& shape constant       & PSF shape Gaussian\\
           & between all 5 images & between all 5 images  & independent of position & in all 5 images\\
           & in image centres     & at any other position & in all 5 images         &\\
\hline
\emph{none/T0006} & \textsf{X} & \textsf{X} & \textsf{X} & \textsf{X}\\
\emph{global}     & $\surd$ & approximately & \textsf{X} & \textsf{X}\\
\emph{local}      & $\surd$ & $\surd$ & $\surd$ & $\surd$\\
\end{tabular}
\label{tab:PSF_schemes}
\end{minipage}
\end{table*}

\section{Catalogue extraction}
\label{sec:catalogues}
Multi-colour catalogues are extracted from a set of PSF homogenised
images in the $ugriz$ filters using SExtractor in dual-image mode. The
procedure is identical to the one presented in
\cite{2009A&A...493.1197E} involving six SExtractor runs. The
unconvolved (i.e. non-PSF-homogenised) $i$-band image is used as the
detection image in all six runs. Five runs are performed with the
convolved (i.e. PSF-homogenised) $ugriz$ images. One additional run
with the unconvolved $i$-band image as detection as well as
measurement image is performed to estimate total $i$-band
magnitudes. This latter run is necessary because the isophotal
apertures, which are used for flux measurements and hence colour
estimation, are defined on the detection image by SExtractor. Using a
measurement image with a larger PSF leads to flux leaking outside the
aperture and therefore underestimated fluxes (overestimated
magnitudes). Those biased fluxes are optimal for colour measurements
because they correspond to the same physical parts of an object (see
Sect.~\ref{sec:PSFhomo} above) but cannot readily be used to estimate
total magnitudes. However, with the total $i$-band magnitudes
estimated reliably in the sixth SExtractor run and the accurate colour
measurements it is still possible to arrive at total magnitude
estimates in all bands, e.g. for a band $X$:
\begin{equation}
X_{\rm tot}=i_{tot}+(X-i)\,,
\end{equation}
where $(X-i)$ is the corresponding colour index.

The multi-colour photometry is complemented with the following
quantities \citep[for details see ][]{2009A&A...493.1197E}:
\begin{itemize}
\item Limiting magnitudes estimated from the local sky-background
  around an object.
\item Extinction values extracted from the \cite{1998ApJ...500..525S}
  maps at the object's position.
\item Masks based on algorithms to reject regions of low
  signal-to-noise ratio (S/N), halos and diffraction spikes of bright
  stars, and asteroids. These automated masks are then inspected by
  eye and further modified if necessary by people in the survey team.
\end{itemize}

\subsection{Photometric quality control and creation of the mosaic catalogue}
\label{sec:quality_control}
Before photometric redshifts are estimated (see
Sect.~\ref{sec:photoz}) several tests are performed with the
multi-colour photometry alone to ensure the integrity of the
  data. The surface densities of objects in a fixed magnitude range
are analysed and the object magnitude number counts are inspected on a
field-by-field basis. The sky distributions of the same objects are
plotted and eye-balled and distributions of quantities like the
half-light-radius or the position-angle of galaxies for each field are
plotted as well as colour-colour diagrams of stars (selected by size
and magnitude).

In order to arrive at a homogeneous mosaic catalogue we define hard
cuts in right ascension and declination for each field. As the
boundary between two fields we choose the mean of the extremal
positions of objects taken from one field and from a neighbouring
field. This method ensures that no celestial object appears more than
once in our mosaic catalogue.\footnote{Note that there could still be
  objects which appear in two neighbouring fields because of
  astrometric errors. But their number is so low that we do not
  account for this here.}

\section{Photometric redshifts}
\label{sec:photoz}
Photo-$z$'s are estimated with the \emph{BPZ} code
\citep{2000ApJ...536..571B,2006AJ....132..926C} one of the most-widely
used photo-$z$ codes. \cite{2010A&A...523A..31H} tested several codes
against simulated and real data showing that \emph{BPZ} is amongst the
most accurate codes when combined with the best available SED
templates.

In the photo-$z$ estimation process we properly take account of
objects that are not detected in one or more of the $ugrz$-bands
(identified by magnitude estimates fainter than the limiting
magnitudes, which can happen when SExtractor is run in dual-image
mode). As in \cite{2009A&A...493.1197E} and \cite{2009A&A...498..725H}
we use the re-calibrated template set of
\cite{2004_Capak_PhDT}.\footnote{This template set is very similar to
  the one used for the T0006 catalogues.} Compared to
\cite{2009A&A...493.1197E} we implemented some changes to improve the
low-redshift behaviour of the photo-$z$'s that are explained in the
following.

\subsection{Modifications of the prior}
\label{sec:prior}
The Bayesian approach used by \emph{BPZ} encompasses the calculation
of the redshift likelihood and its subsequent multiplication with a
prior to yield the posterior probability of an object having a certain
redshift given the data. For noisy data the peak in the likelihood
always has a finite width. For a $z=0$ object the following happens:
the likelihood function, which extends to redshifts $z>0$ because of
its finite width, is multiplied with a steep prior that behaves like
$P(z)\approx z^{\alpha_t}$ for $z\ll1$, with $0.9\la \alpha_t\la
2.5$. In particular $P(z=0)=0$ so that the peak of the posterior
probability distribution is always at $z>0$. Since \emph{BPZ} picks
the peak of the posterior as the photo-$z$ estimate, this leads to a
systematic over-estimation of photo-$z$'s at low redshift, whenever
there is an appreciable number of low-$z$ objects with limited S/N
(and hence broad likelihoods) to make this effect visible. Similar
biases of different severity can be seen in the Bayesian photo-$z$'s
of e.g. \citeauthor{2003AJ....125..580C}
(\citeyear{2003AJ....125..580C}; Fig.~7),
\citeauthor{2000ApJ...536..571B} (\citeyear{2000ApJ...536..571B};
Fig.~7), and \citeauthor{2009A&A...500..981C}
(\citeyear{2009A&A...500..981C}; Figs.~3~\&~5). This behaviour is a
general problem of template-based, Bayesian redshifts. Here we present
an ad-hoc solution, but emphasise that further research is needed on
this.

\begin{figure}
\centering
\includegraphics[width=0.7\hsize]{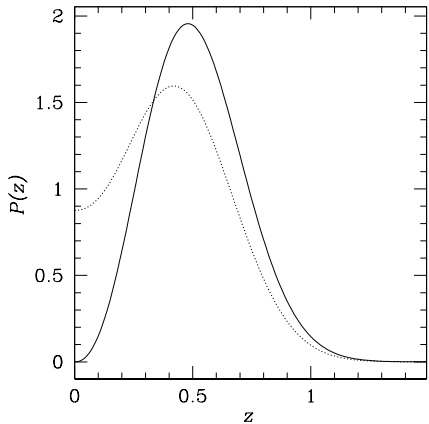}
\caption{Redshift prior before (solid) and after (dotted) modification, here shown for an elliptical galaxy with $i=20$.}
\label{fig:prior}
\end{figure}

To attenuate low-redshift bias we run \emph{BPZ} twice, first with the
original prior from \cite{2000ApJ...536..571B}
\begin{equation}
P(z)\propto z^{\alpha_t}\,\exp{\left[-\left(\frac{z}{z_{mt}(m_0)}\right)^{\alpha_t}\right]}
\end{equation}
and again with a modified prior
\begin{equation}
\label{eq:prior_mod}
P_{\rm mod}(z)\propto \left(z^{\alpha_t}+{\bf
  0.05}\right)\,\exp{\left[-\left(\frac{z}{z_{mt}(m_0)}\right)^{\alpha_t}\right]}\,.
\end{equation}
This modified prior no longer vanishes for $z=0$, but levels
off. Figure~\ref{fig:prior} illustrates this modification. We keep the
result of the first run unless the most probable redshift is below
$z=0.1$ and the ODDS parameter \cite[i.e. the fraction of the
  integrated probability included in the primary peak of the posterior
  probability distribution; see][]{2000ApJ...536..571B} associated
with this solution is smaller than 0.8 indicating a possibly biased
redshift. We found that this ad-hoc modification yields an improved
low-redshift performance when comparing photo-$z$'s to spectroscopic
redshifts (spec-$z$'s; see next section). The choice of adding 0.05 in
the first term of Eq.~\ref{eq:prior_mod} yields the best results for
this particular data set. But data with different noise properties
might require different modifications. As mentioned above these prior
modifications are a subject of ongoing research. It will be important
to find robust, self-consistent ways of finding the optimal
modification, possibly based on a large, spectroscopic, low-$z$
training set.

\subsection{Photo-$z$ accuracy}
\label{sec:accuracy}
\begin{figure*}
{\it none\hspace{5cm}global\hspace{5.2cm}local\\}
\includegraphics[width=0.32\hsize]{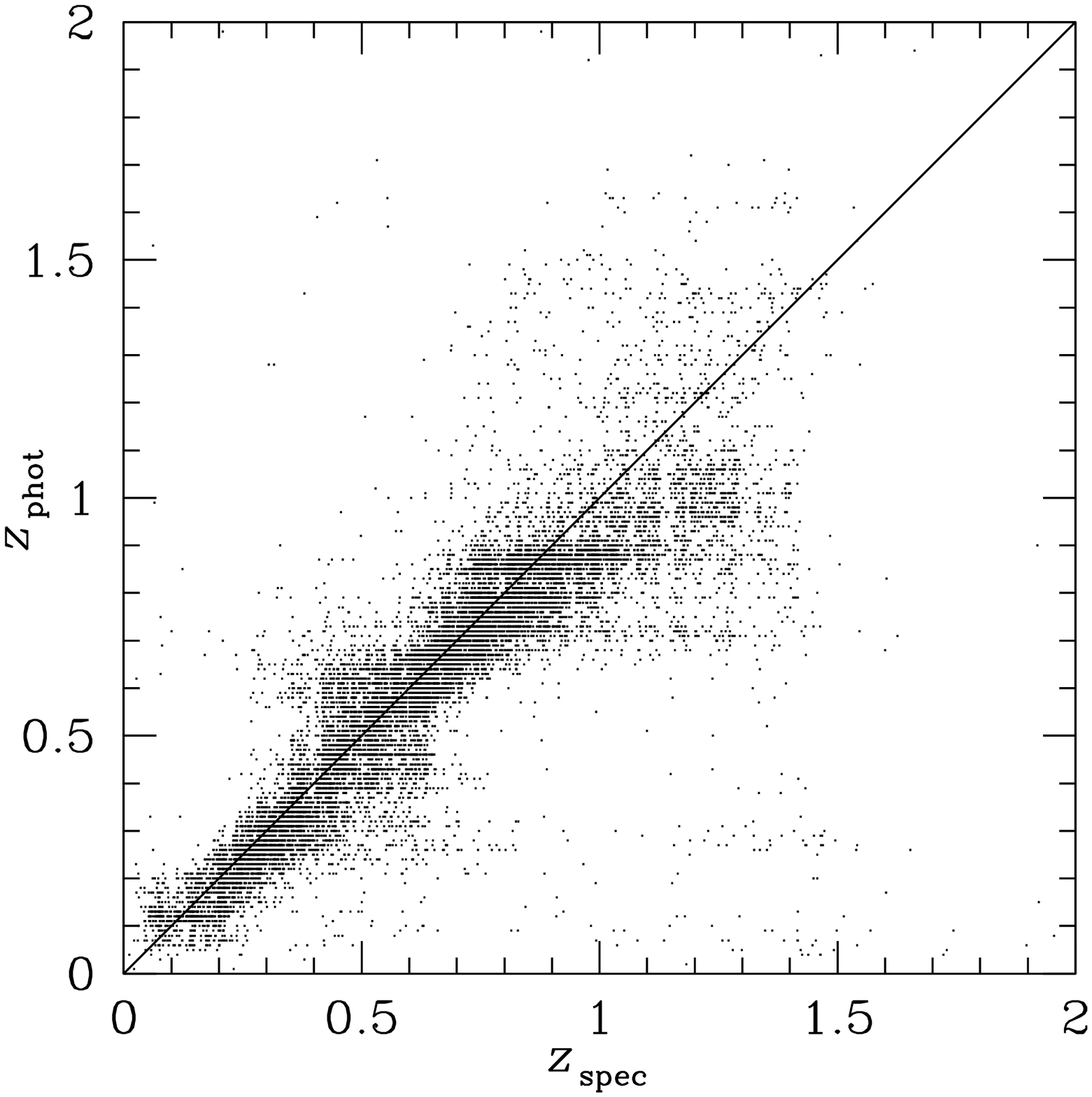}
\includegraphics[width=0.32\hsize]{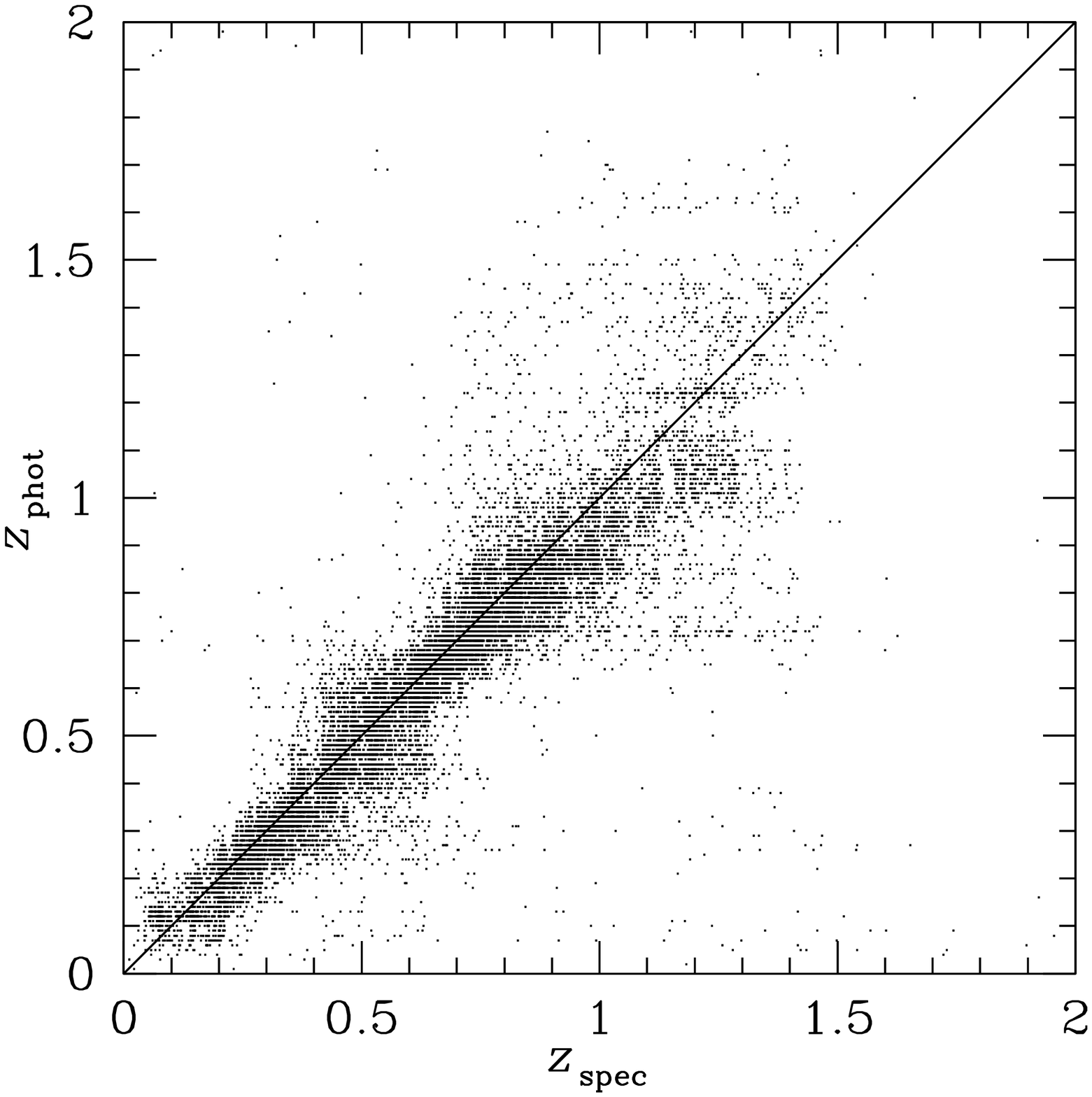}
\includegraphics[width=0.32\hsize]{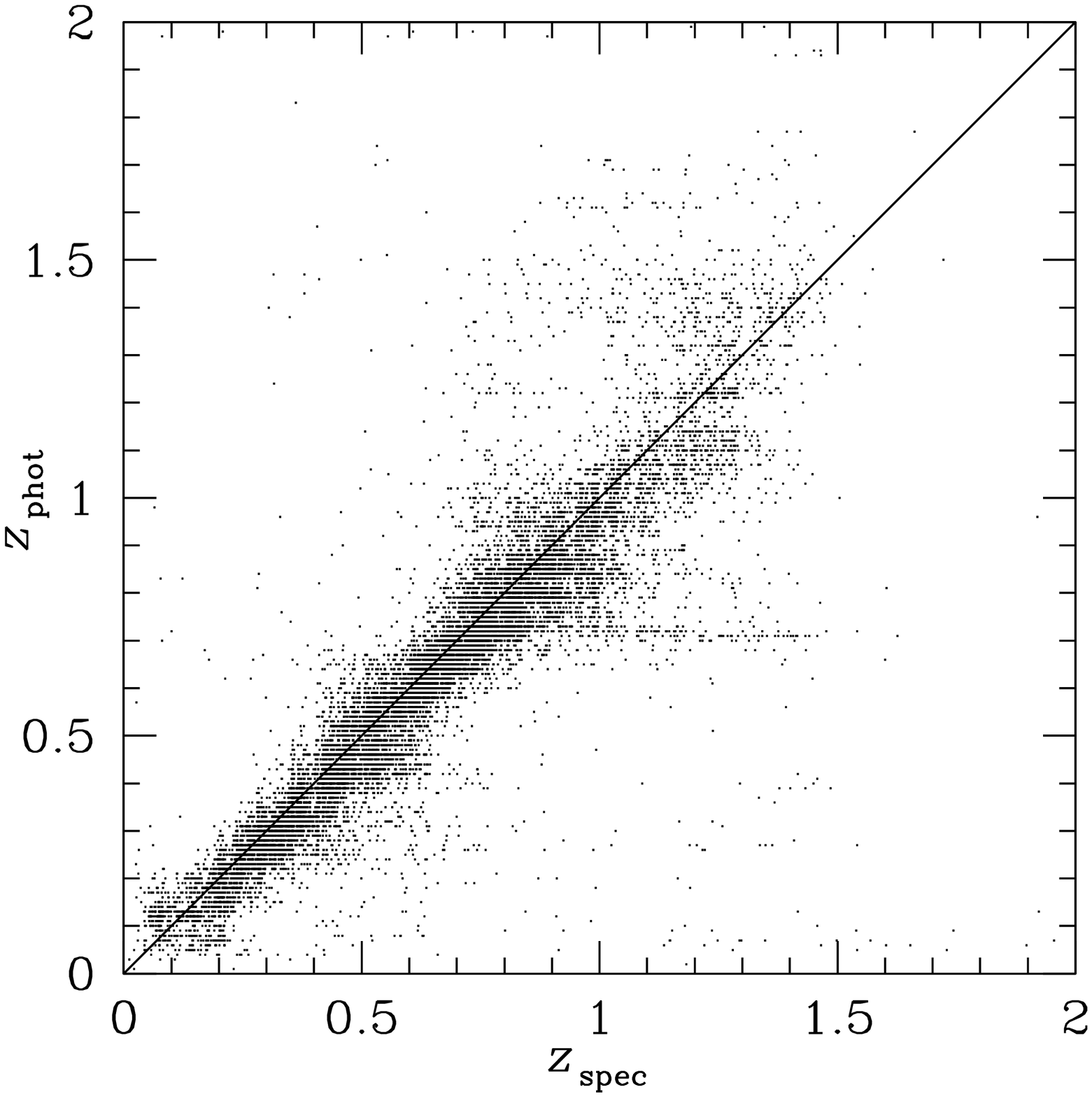}
\caption{Photo-$z$'s vs. spec-$z$'s for the three different ways of
  homogenising the PSF. Shown are all objects with secure spec-$z$'s
  (from VVDS, DEEP2, and SDSS) of the 20 fields with VVDS or DEEP2
  overlap. No magnitude or ODDS cut is applied.}
\label{fig:zz}
\end{figure*}

We estimate the accuracy of the CFHTLenS photo-$z$'s by comparing them
to spec-$z$'s from the VIMOS VLT Deep Survey \citep[VVDS;
][]{2005A&A...439..845L,2008A&A...486..683G} and the DEEP2 galaxy
redshift survey \citep{2007ApJ...660L...1D} which overlap with 20
fields of the CFHTLS-Wide. The deepest spectroscopic fields contain
objects down to $i=24$, however with an increasing incompleteness for
fainter magnitudes. We also add spec-$z$'s from the Sloan Digital Sky
Survey seventh data release \citep[SDSS-DR7; ][]{2009ApJS..182..543A}
if available in those 20 fields. We do not add the SDSS spec-$z$'s in
the other $\sim90$ CFHTLS-Wide fields that overlap with SDSS because
they do not contain significant additional information due to their
low redshift. In Fig.~\ref{fig:zz} the photo-$z$'s for the three
different PSF homogenisation methods are shown against the spec-$z$'s.
Visibly the photo-$z$ accuracy improves with increasing sophistication
of the PSF homogenisation.

We limit the analysis to galaxies, as identified by the spec-$z$
surveys. The stellar spectra will be used in
Sect.~\ref{sec:star_galaxy} to develop criteria for star-galaxy
separation. It should be noted that the spec-$z$ catalogues used here
are incomplete for $i\ga22$ and might paint too positive a picture of
the photo-$z$ accuracy \citep[see ][for extensive discussions of these
  effects]{2008A&A...480..703H,2010A&A...523A..31H}. In
Sect.~\ref{sec:conclusions} we discuss strategies to acquire a
complete picture of the photo-$z$ accuracy that will be presented in
Benjamin et al. (in prep.).

For each object with a reliable spec-$z$ measurement (quality flags 3,
4, 23, 24 for VVDS; quality flags 3 and 4 for DEEP2; quality flag 3
for SDSS) we calculate the quantity $\Delta z = \frac{z_{\rm
    phot}-z_{\rm spec}}{1+z_{\rm spec}}$, where $z_{\rm phot}$ is the
peak of the posterior probability distribution. Objects with
$\left|\Delta z\right| >0.15$ are regarded as outliers.\footnote{This
  choice is arbitrary and is mainly used for historical reasons since
  many photo-$z$ studies in the past have adopted the same definition
  of an outlier. With a typical photo-$z$ scatter of $\sigma\sim0.04$
  it ensures that only $\approx4$-$\sigma$ outliers are counted.} For
a given sample we then calculate the mean of $\Delta z$ and the
standard deviation around this mean, which corresponds to the
photo-$z$ scatter. This is done after outliers have been
excluded. Furthermore we report the total bias of the sample, i.e. the
mean of $\Delta z$ including the outliers. We would like to stress
that these three numbers (outlier fraction, scatter, bias) are not
independent of each other and cannot reflect the full error
distribution, which is highly non-Gaussian.

These statistics are calculated for different narrow $i$-band
magnitude bins as well as for different narrow redshift bins (with a
pre-selection of $19<i<24.5$) for each of the three different PSF
homogenisation approaches, \emph{none}, \emph{global}, and
\emph{local}, as well as for the \emph{T0006} catalogues. Errors are
calculated assuming Poissonian shot noise. It should be noted that
there is non-negligible correlation between the errors in neighbouring
magnitude/redshift bins.

\begin{figure*}
\includegraphics[width=0.45\textwidth]{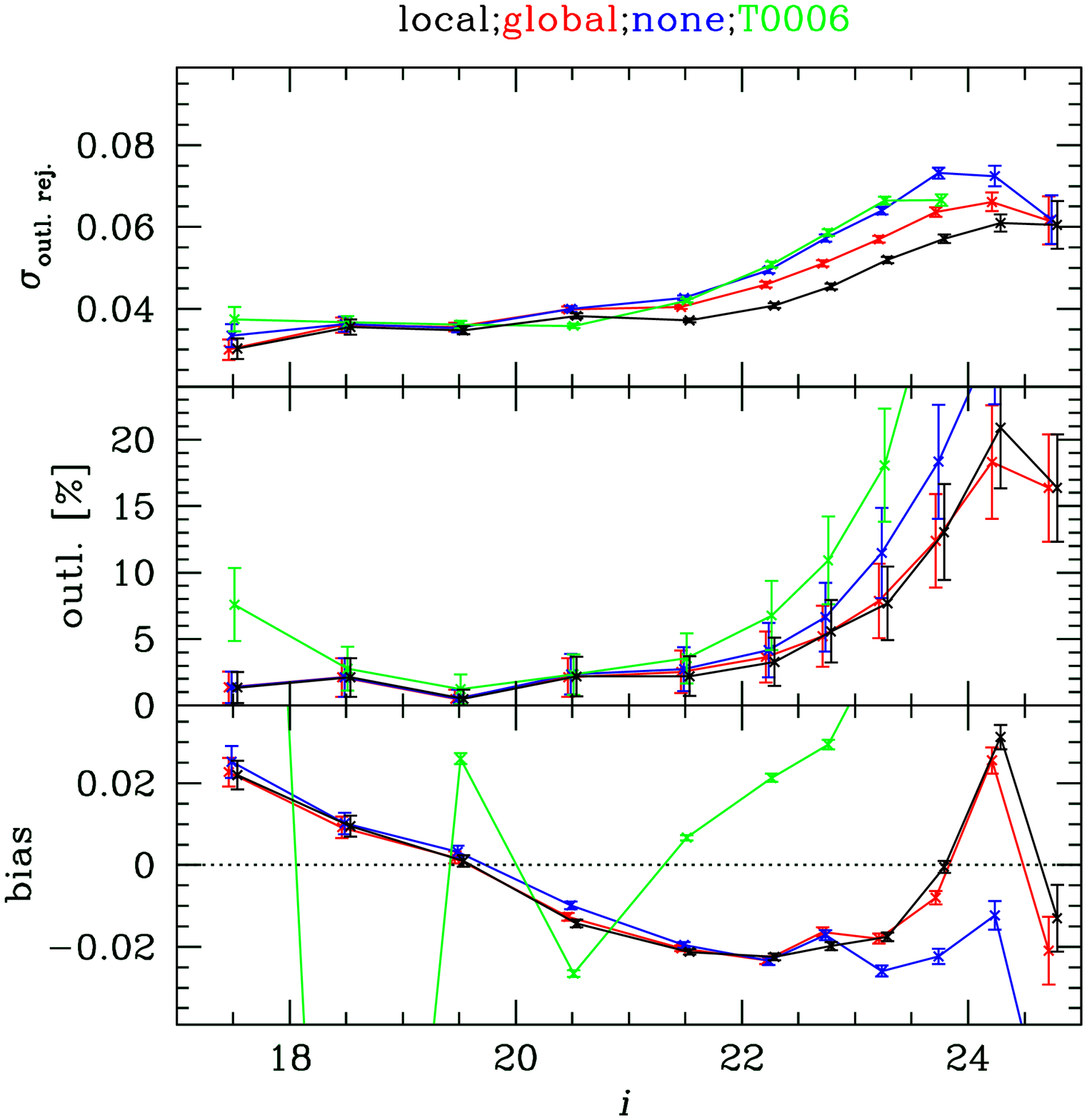}
\includegraphics[width=0.45\textwidth]{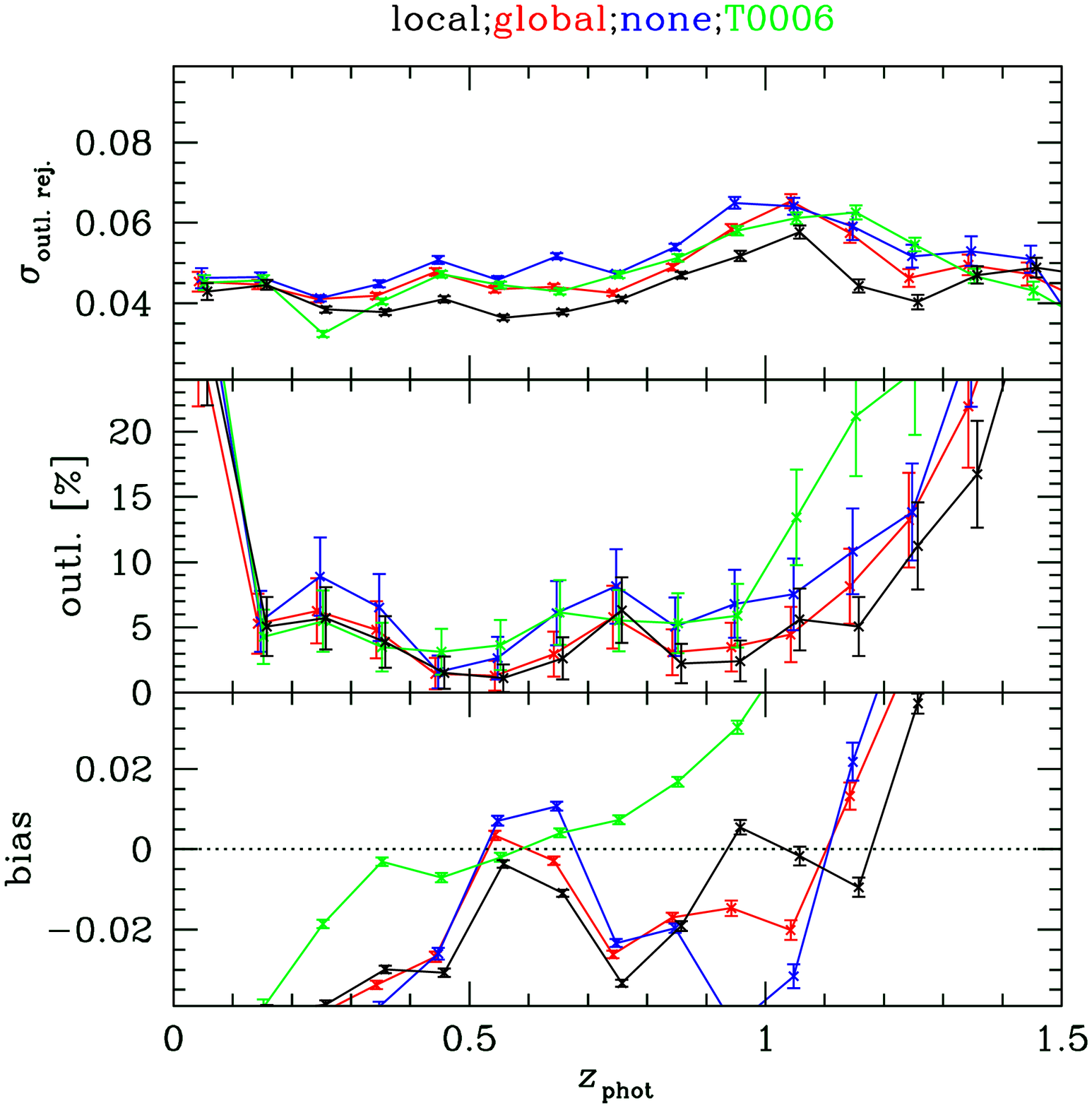}
\caption{Photo-$z$ statistics as a function of magnitude (left) and
  redshift (right) showing the effects of different ways of
  homogenising the PSF. While \emph{none} and \emph{T0006} correspond
  to no PSF homogenisation, \emph{global} corresponds to a constant
  Gaussian kernel used for the convolution of the image in one band,
  and \emph{local} corresponds to a non-Gaussian, spatially varying
  kernel that leads to the same Gaussian PSF over the whole field. The
  top panel shows the photo-$z$ scatter after outliers were rejected,
  the middle panel shows the outlier rate, and the bottom panel shows
  the bias (outliers included; positive means photo-$z$'s overestimate
  the spec-$z$'s). Errors are purely Poissonian. Note that the
    errors between magnitude/redshift bins are correlated.}
\label{fig:z_phot_stats_PSFhomo}
\end{figure*}

Figure~\ref{fig:z_phot_stats_PSFhomo} shows the photo-$z$ accuracy as
a function of $i$-band magnitude and redshift for the different
methods. The effects of the PSF homogenisation can be clearly
seen. While the performance at bright magnitudes is similar in all
methods the photo-$z$ scatter and outlier rates at fainter magnitudes
for the two methods without PSF homogenisation (\emph{none} and
\emph{T0006}) are larger than for the methods with PSF homogenisation
(\emph{global} and \emph{local}). Fainter objects are also smaller on
average and hence their shape is more strongly dominated by the
PSF. Not correcting for PSF-effects biases their colours and leads to
less accurate photo-$z$'s.

Looking at the accuracy as a function of redshift shows that PSF
homogenisation leads to greater accuracy over the whole redshift
range. The effect is more pronounced at higher redshifts, but since
there are also many faint low-redshift galaxies the low-$z$ statistics
for \emph{global} and \emph{local} are generally better than for \emph{none}
and \emph{T0006} as well.

The \emph{global} and \emph{local} schemes show similar redshift accuracy. At
redshifts $z\ga0.9$ \emph{local} shows somewhat reduced photo-$z$ scatter,
but the differences are small. This finding supports the hypothesis
from Sect.~\ref{sec:PSFhomo_general} that although the PSF still
varies from centre to edge in the \emph{global} images (unlike the \emph{local}
images) these variations are similar in all bands and do not lead to
strong colour biases.

The photo-$z$ bias is non-negligible for most magnitudes and redshifts
regardless of the method. With typical template-based photo-$z$
methods it is very hard to suppress this bias without introducing
larger scatter or more outliers. The strategy for scientific studies
using such photo-$z$'s must be to properly calibrate and account for
this bias in the analysis. We do not correct for the bias at this
stage.

\begin{figure*}
\includegraphics[width=0.45\textwidth]{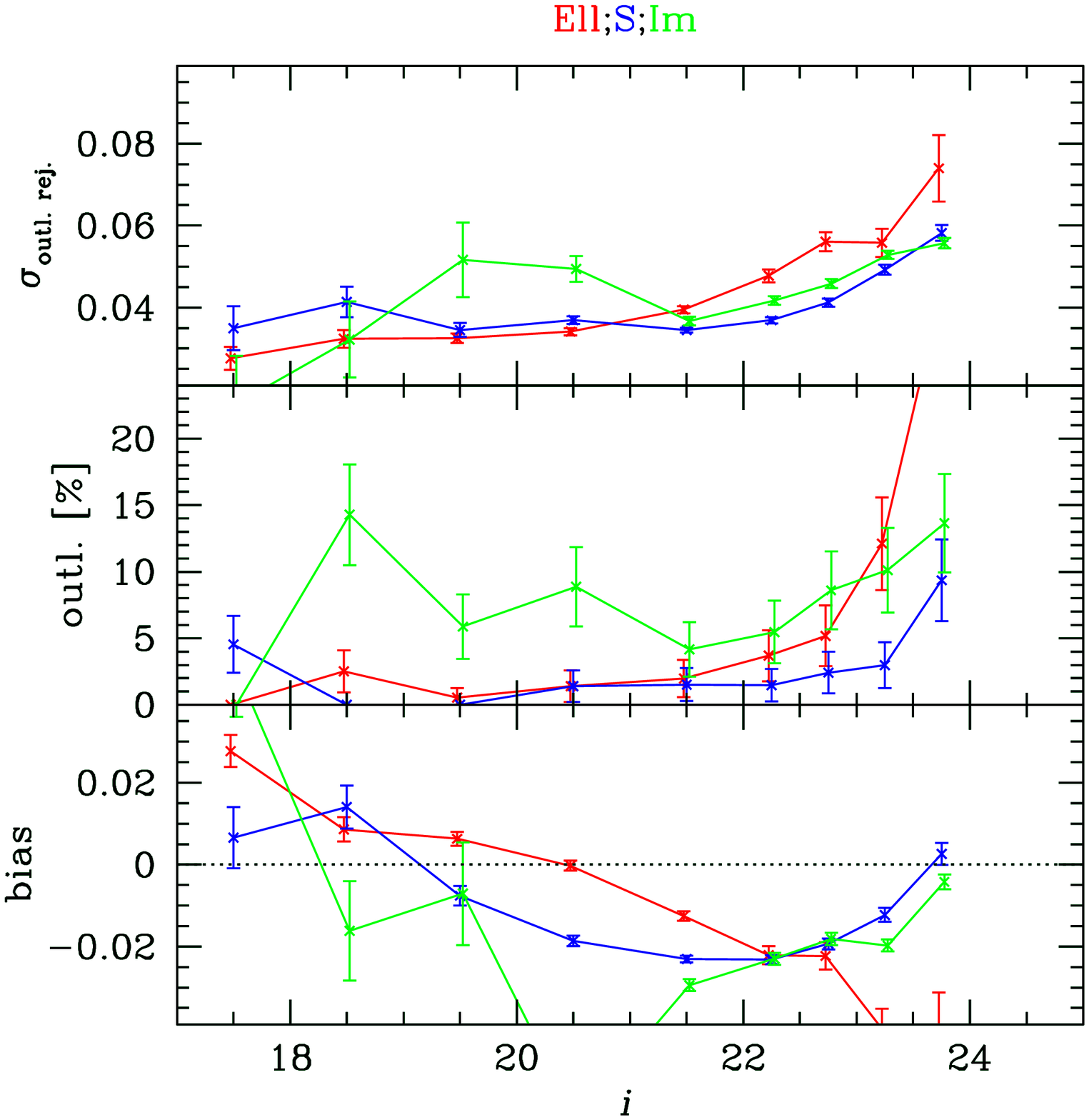}
\includegraphics[width=0.45\textwidth]{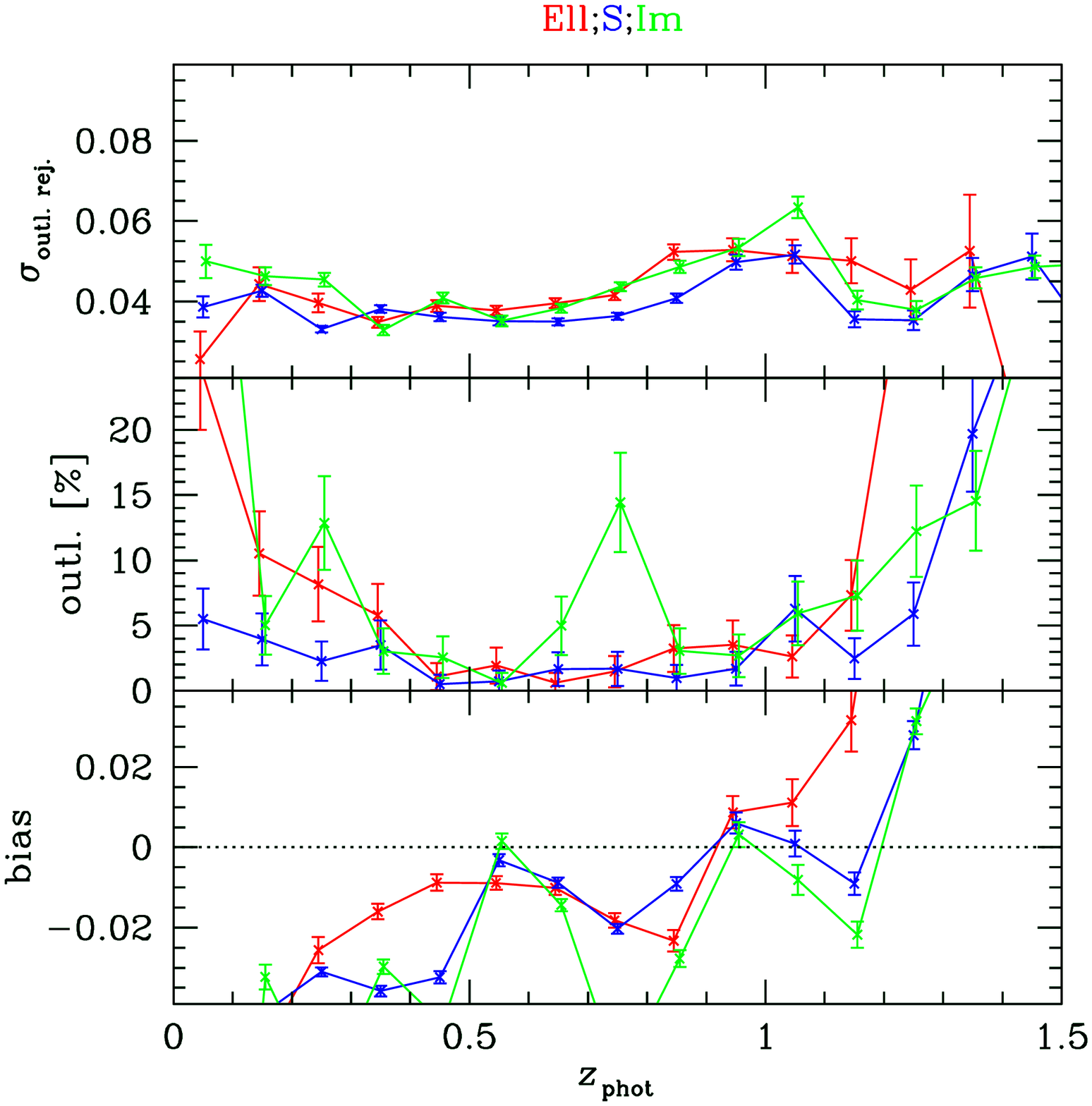}
\caption{Same as Fig.~\ref{fig:z_phot_stats_PSFhomo} but showing the
  accuracy for the different SED templates using the \emph{local}
  photometry.}
\label{fig:z_phot_stats_type}
\end{figure*}

In Fig.~\ref{fig:z_phot_stats_type} we present the photo-$z$ accuracy
for the different SED templates as determined by the photo-$z$
code. Similar to \cite{2006A&A...457..841I} we find that the accuracy
for elliptical galaxies and spiral galaxies is very similar. Only
actively star-forming galaxies show a degraded photo-$z$ accuracy. It
is obvious that the accuracy for ellipticals/spirals suffers
considerably once the $4000\AA$-/Balmer-break starts to leave the
filter set at $z\approx1.1$/$1.3$.

\subsection{Re-calibration of the photometric zeropoints}
\label{sec:recalib}
It has been suggested in the literature
\citep[e.g.][]{2006AJ....132..926C,2006A&A...457..841I,2009A&A...500..981C}
that a re-calibration of the photometric zeropoints of the images with
the help of spec-$z$'s can lead to an enhanced accuracy of
photo-$z$'s. The procedure involves running the photo-$z$ code on a
sub-sample of objects with reliable spec-$z$'s and just fitting for
the template while fixing the redshift to the spectroscopic value. The
averaged magnitude differences in a band between the best-fitting
templates and the observed photometry can then be applied as
corrections to the zeropoint in that band. Usually this is done
iteratively until convergence is reached.

\begin{figure}
\centering
\includegraphics[width=\hsize]{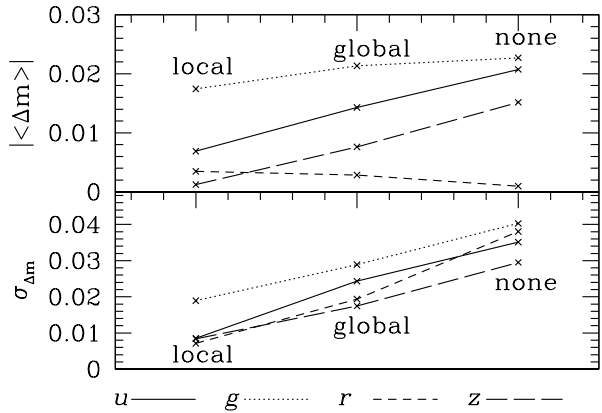}
\caption{Absolute values of the means (top) and standard deviations (bottom) of the 20 zeropoint offsets for the different PSF homogenisation methods and different bands.}
\label{fig:offsets_mean}
\end{figure}

We performed such a zeropoint re-calibration for the \emph{none},
\emph{global}, and \emph{local} methods in the 20 fields with VVDS/DEEP2
coverage. This yields 20 zeropoint corrections for four of the five
bands\footnote{The re-calibration procedure is only sensitive to
  colours. So we decide to fix the offset in the $i$-band - the
  detection band - to $\Delta i=0$.} and for each method. The absolute
value of the mean and the standard deviation of the 20 zeropoint
offsets are shown in
Fig.~\ref{fig:offsets_mean}.

From the figure it is clear that both the mean and the width of the
distributions become smaller going from \emph{none} to \emph{global} to
\emph{local}. In particular, the corrections for the \emph{local} method
mostly vanish, i.e. most \emph{local} offsets are similar to or smaller
than the error of a single correction ($\approx0.02$mag).

The zeropoints used for the catalogue extraction were identical in all
three methods. Thus, our results for the zeropoint re-calibrations
strongly suggest that mostly PSF-effects are corrected by such a
procedure.\footnote{Certainly, there are other effects that play a
  role here. For example, absolute photometric calibrations are often
  done with standard stars that were observed in a slightly different
  photometric system than the instrumental one. Conversions between
  the standard system and the instrumental system depend on the colour
  term of the object. Thus, a correction that is correct on average
  for stars is not correct on average anymore for galaxies, which have
  very different spectral energy distributions. Often also the
  filter-curves used for template-based photo-$z$'s have some
  uncertainty, especially in the ultra-violet where the differential
  atmospheric transparency has a large influence on the effective
  throughput.} Furthermore it means that proper PSF homogenisation, in
combination with an accurate absolute photometric calibration,
supersedes zeropoint re-calibration procedures like the ones presented
in \cite{2006A&A...457..841I,2009A&A...500..981C}.

\begin{figure*}
\includegraphics[width=0.45\textwidth]{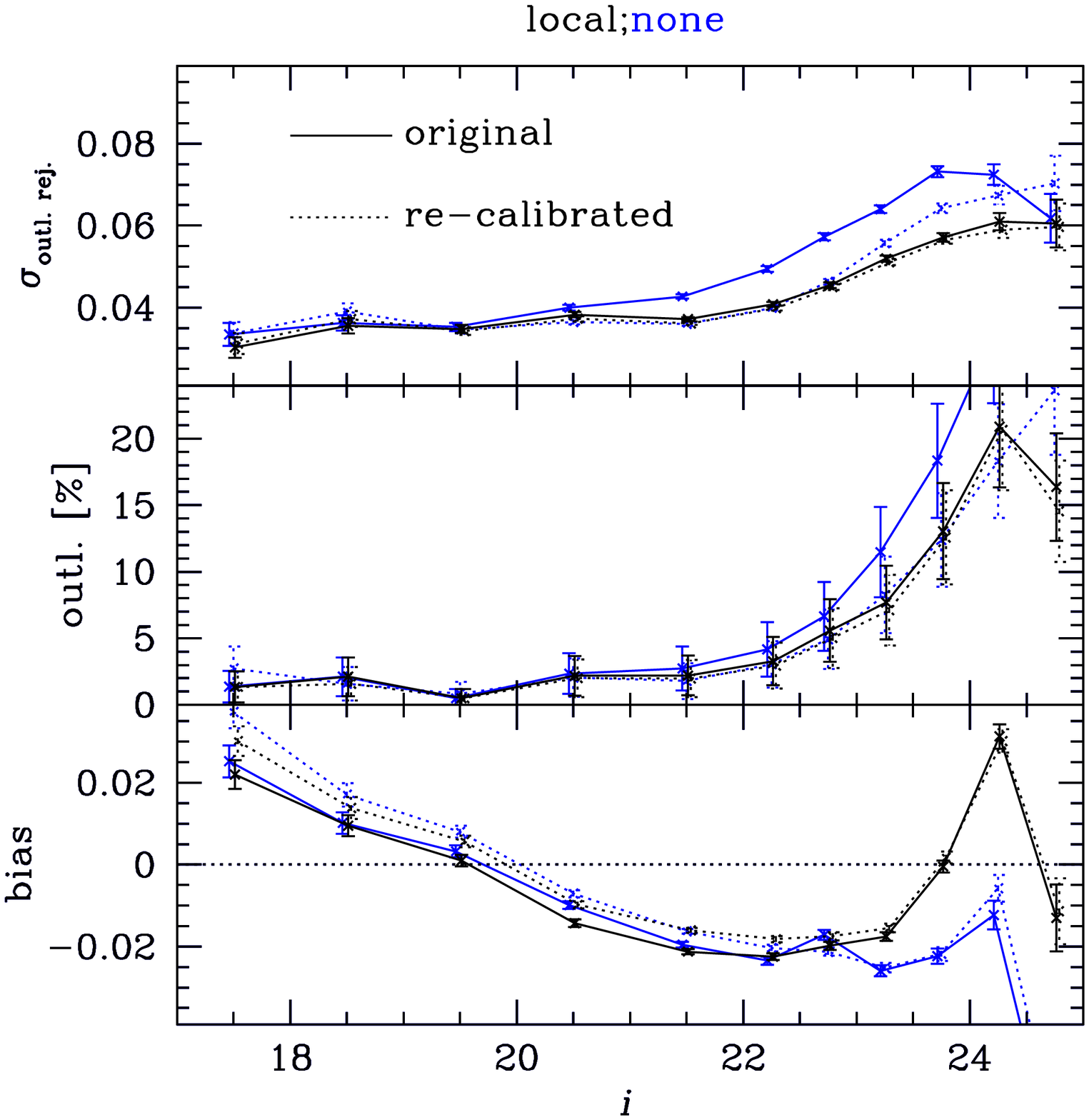}
\includegraphics[width=0.45\textwidth]{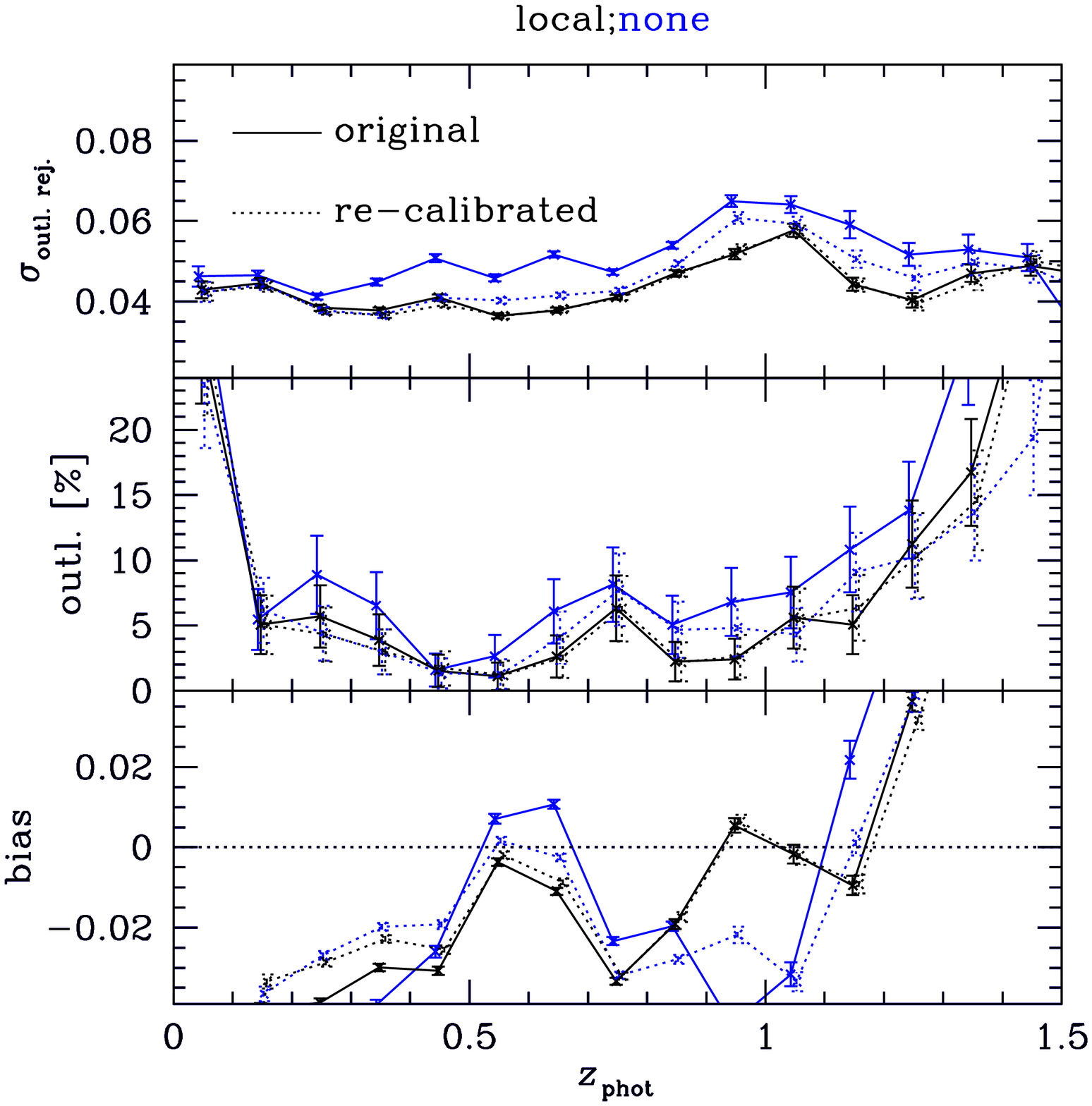}
\caption{Same as Fig.~\ref{fig:z_phot_stats_PSFhomo} but showing the
  effects of zero point re-calibration on the \emph{none} method
  (i.e. without PSF-homogenisation) in comparison to the default
  \emph{local} method (i.e. local PSF Gaussianisation). The solid
  lines correspond to the original methods whereas the dotted lines
  correspond to the re-calibrated methods. It is clearly visible that
  the re-calibrated \emph{none} method performs similar to the
  original (i.e. not re-calibrated) \emph{local} method, suggesting
  strongly that zeropoint re-calibrations are mostly correcting for
  PSF-effects and not for real biases in the photometric zeropoints.}
\label{fig:z_phot_stats_recalib}
\end{figure*}

This is confirmed by looking at the photo-$z$ accuracy before and
after re-calibration. In Fig.~\ref{fig:z_phot_stats_recalib} the
photo-$z$ statistics are shown again as a function of magnitude and
redshift, for the \emph{none} method and a re-calibrated \emph{none}
method. The \emph{local} method is also plotted as a benchmark. The
improvement is striking although the accuracy of the 'none-recalib'
method does not reach the accuracy of the \emph{local} method at the
faintest magnitudes. Interestingly the lines for a re-calibrated
\emph{local} method are nearly indistinguishable from the basic \emph{local}
method.

We would like to stress that these findings do not tell the whole
story and the situation might be even worse for re-calibrated
photometry. Re-calibrations of the photometric zeropoints might well
fool the user into believing that the photo-$z$ accuracy is better
than it is in reality. If most of the corrections are due to
PSF-effects as our results suggest, then these corrections depend on
the average angular size of the objects. But the average size of the
photometric galaxy sample used for the science projects often differs
from the average size of the spectroscopic calibration sample which is
not only used for the zeropoint re-calibration but also for the
following assessment of the photo-$z$ accuracy. This circular use is
dangerous if the systematic effects that are corrected for depend on
the nature of the objects (e.g. their size) and are not identical for
all objects (like e.g. real photometric zeropoints). Furthermore, if
the zeropoint corrections depend sensitively on seeing it is not
advisable to apply the correction found on one particular field to
another field.

Based on these findings we decide to use the \emph{local} method for all
scientific projects in the CFHTLenS. It offers the best photo-$z$
accuracy combined with the most stable photometry and does not require
re-calibration of the zeropoints to achieve this. The photo-$z$'s for
this method are well understood in the redshift range $0.1<z_{\rm
  phot}<1.3$ for $i<24.5$ - two magnitudes fainter than the analysis
presented in \cite{2009A&A...500..981C} - with photo-$z$ scatter
values in the range $0.03<\sigma<0.06$ and outlier rates smaller than
10 per cent.

\subsection{Selection of sub-samples with higher photo-$z$ accuracy}
\label{sec:ODDS}
\begin{figure*}
\includegraphics[width=0.45\textwidth]{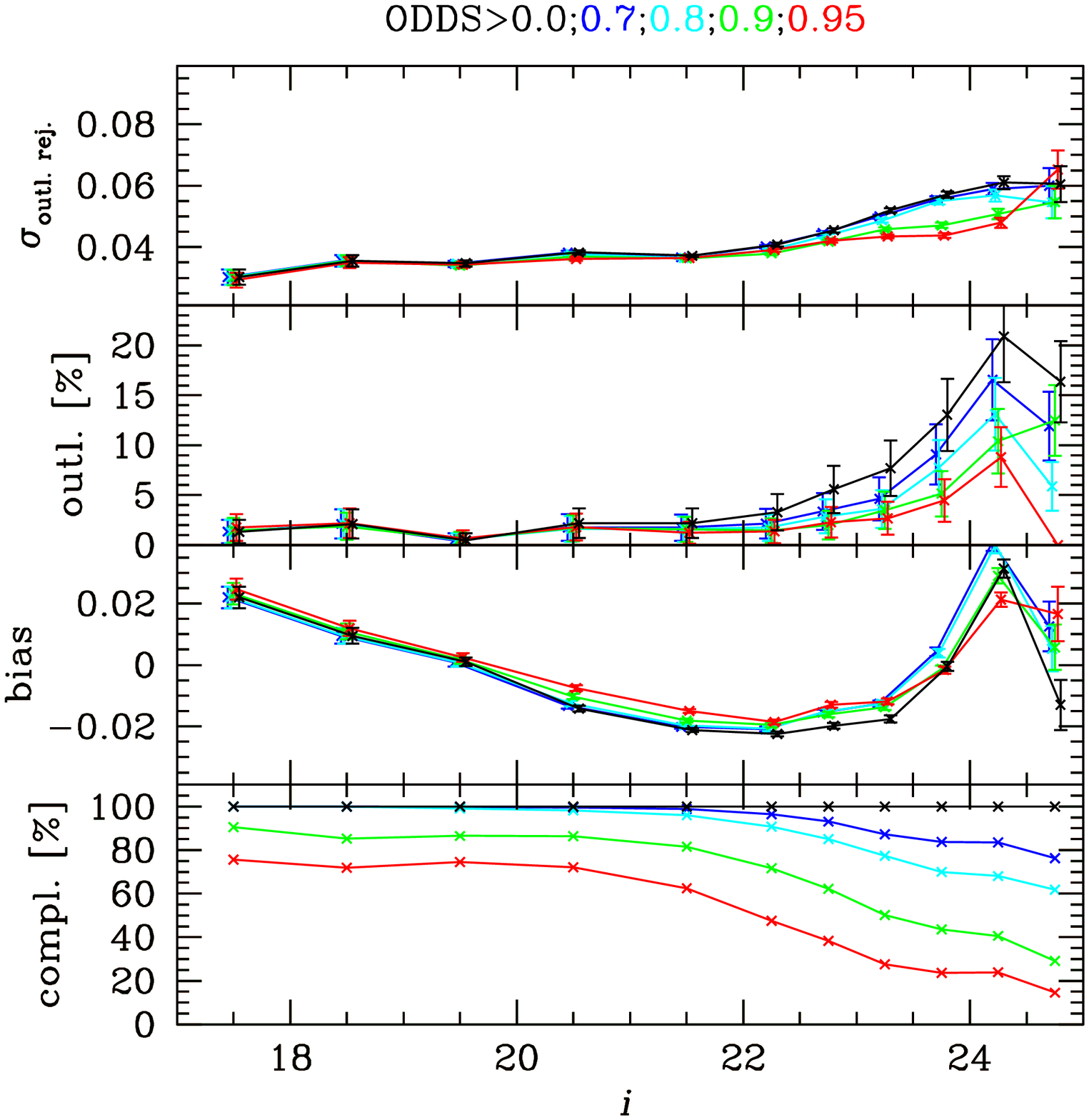}
\includegraphics[width=0.45\textwidth]{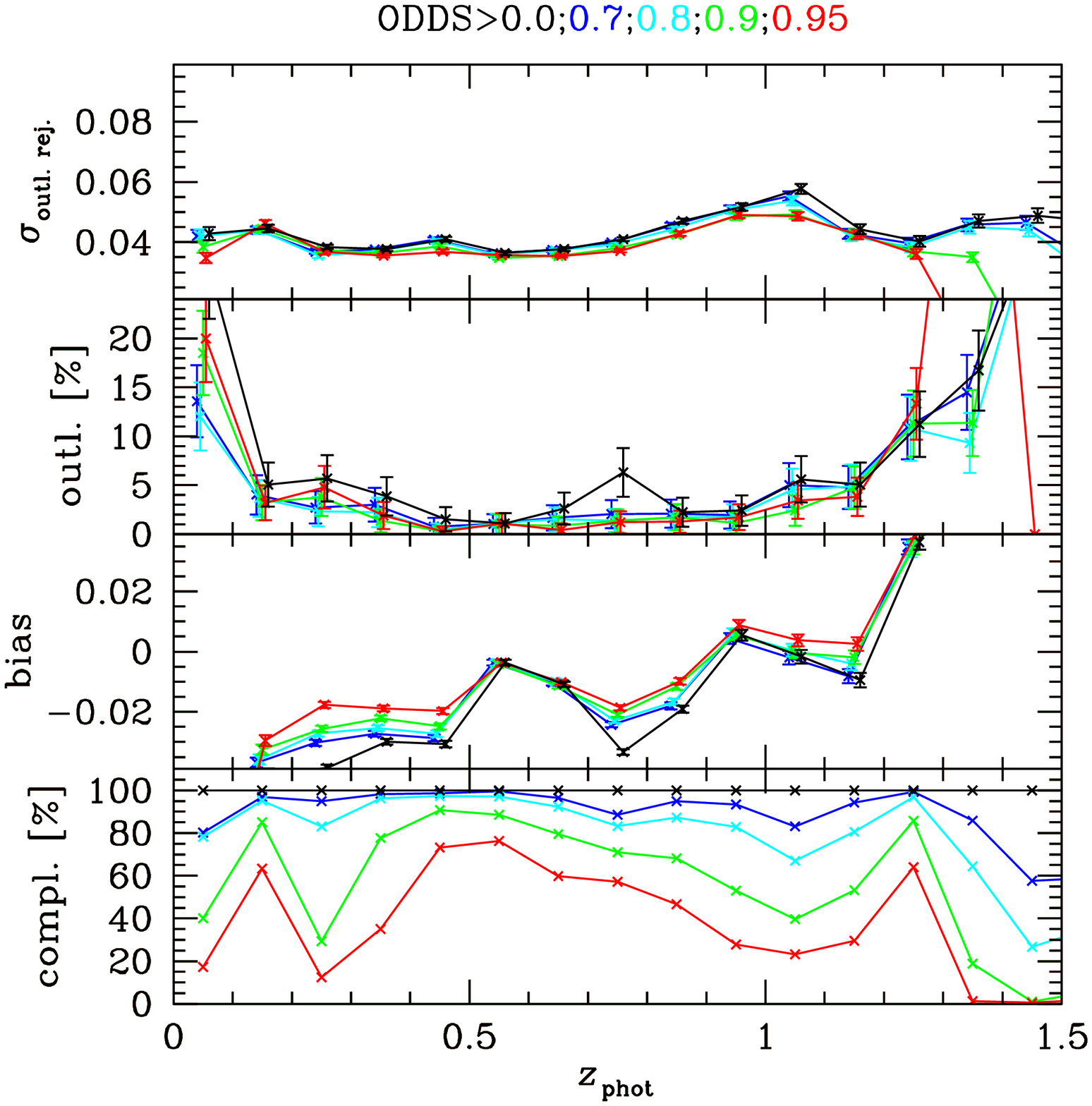}
\caption{Same as Fig.~\ref{fig:z_phot_stats_PSFhomo} but showing the
  effect of cutting on ODDS for the default \emph{local} method
  (i.e. local PSF Gaussianisation; no zero point re-calibration). The
  additional bottom panel shows the completeness of the galaxy sample
  after the ODDS cut is applied.}
\label{fig:z_phot_stats_ODDS}
\end{figure*}

The ODDS parameter introduced by \cite{2000ApJ...536..571B} and
described in Sect.~\ref{sec:prior} can be used to select sub-samples
of galaxies with a higher photo-$z$ accuracy, with the trade-off
  of a decreased completeness and an implicit colour selection. In
Fig.~\ref{fig:z_phot_stats_ODDS} the photo-$z$ statistics for
different cuts on the ODDS parameter are shown. Also the completeness
of the sample is reported when such cuts are applied.

The main effect of a cut on ODDS is that the outlier rates at faint
magnitudes are reduced. However, by looking at the redshift dependence
(right panel of Fig.~\ref{fig:z_phot_stats_ODDS}) it becomes clear
that these problematic objects are mostly assigned very high redshifts
of $z_{\rm phot}>1.5$ (i.e. they don't appear in the right-hand
panel). One exception is a feature at $z_{\rm phot}\sim0.75$ where the
outlier rate can be effectively suppressed by cutting on ODDS. But
overall ODDS has a negligible impact in the well-understood redshift
range of $0.1<z_{\rm phot}<1.3$. For most applications it is probably
not meaningful to apply a global cut on ODDS, most importantly because
such a selection always entails an implicit colour selection. A
redshift-dependent cut on ODDS could make sense for applications that
could tolerate such a selection (e.g. selection of background sources
for weak gravitational lensing).

\subsection{Star-galaxy separation}
\label{sec:star_galaxy}
We separate stars from galaxies using a combination of size, magnitude
and colour information.  Although a pure stellar sample can be
obtained by isolating the stellar branch in the size-magnitude plane,
the mixing of faint and small galaxies with stars considerably
complicates the separation of both classes of objects in this
regime. For science cases requiring a pure and complete galaxy sample,
having a robust star-galaxy estimator becomes a key issue.

A great advantage of template-fitting methods to estimate photo-$z$'s
is the ability to use different template sets. Therefore, in addition
to galaxy templates, one is able to test for stellar templates as
well. We run \emph{BPZ} again fixing the redshift to $z=0$ and using
the stellar spectral library from \cite{1998PASP..110..863P}. For each
object, \emph{BPZ} outputs a best-fitting estimate from the stellar
library with an associated $\chi^2_{star}$ which can be compared to
$\chi^2_{gal}$ associated with the best-fitting galaxy template and
redshift. Then ideally a star would satisfy the relation
$\chi^2_{star} < \chi^2_{gal}$.  However, one should keep in mind that
both estimators were computed from independent template libraries and
the comparison is therefore not straightforward because of different
degrees of freedom. To overcome this difficulty, the estimator can be
tested and calibrated on spectroscopic data. The method was applied
previously on the CFHTLS Wide \citep{2009A&A...500..981C} and tested
using spectroscopic data from VVDS F02 \citep{2005A&A...439..845L} and
VVDS F22 \citep{2008A&A...486..683G} surveys. The following criteria,
combining size, magnitude and colour information, have been found to
give the best compromise between a pure and a complete galaxy sample
when tested on the VVDS:
\begin{itemize}
\item for $i < 21$: all objects with $r_h < r_{h,{\rm limit}}$ are flagged as stars,
\item in the range $21 < i < 23$, objects with $r_h < r_{h,{\rm limit}}$ \emph{and}
 $\chi^2_{star} <  2\times \chi^2_{gal}$ are flagged stars,
 \item and for  $i > 23$ all objects are flagged as galaxies,
\end{itemize}
where $r_{h}$ is the average half light radius in a field computed in
the $i$-band image and $r_{h,{\rm limit}}$ the 3-$\sigma$ upper limit
of the $r_{h}$ distribution in a single image. The $r_{h,{\rm limit}}$
values for each field are determined by manually inspecting the
size-magnitude diagrams.\footnote{It should be noted that the $r_h$
  estimation becomes less and less reliable at faint magnitudes. But
  for $i<23$, where we use it, it is still largely unbiased.}

\begin{figure*}
\includegraphics[width=0.45\textwidth]{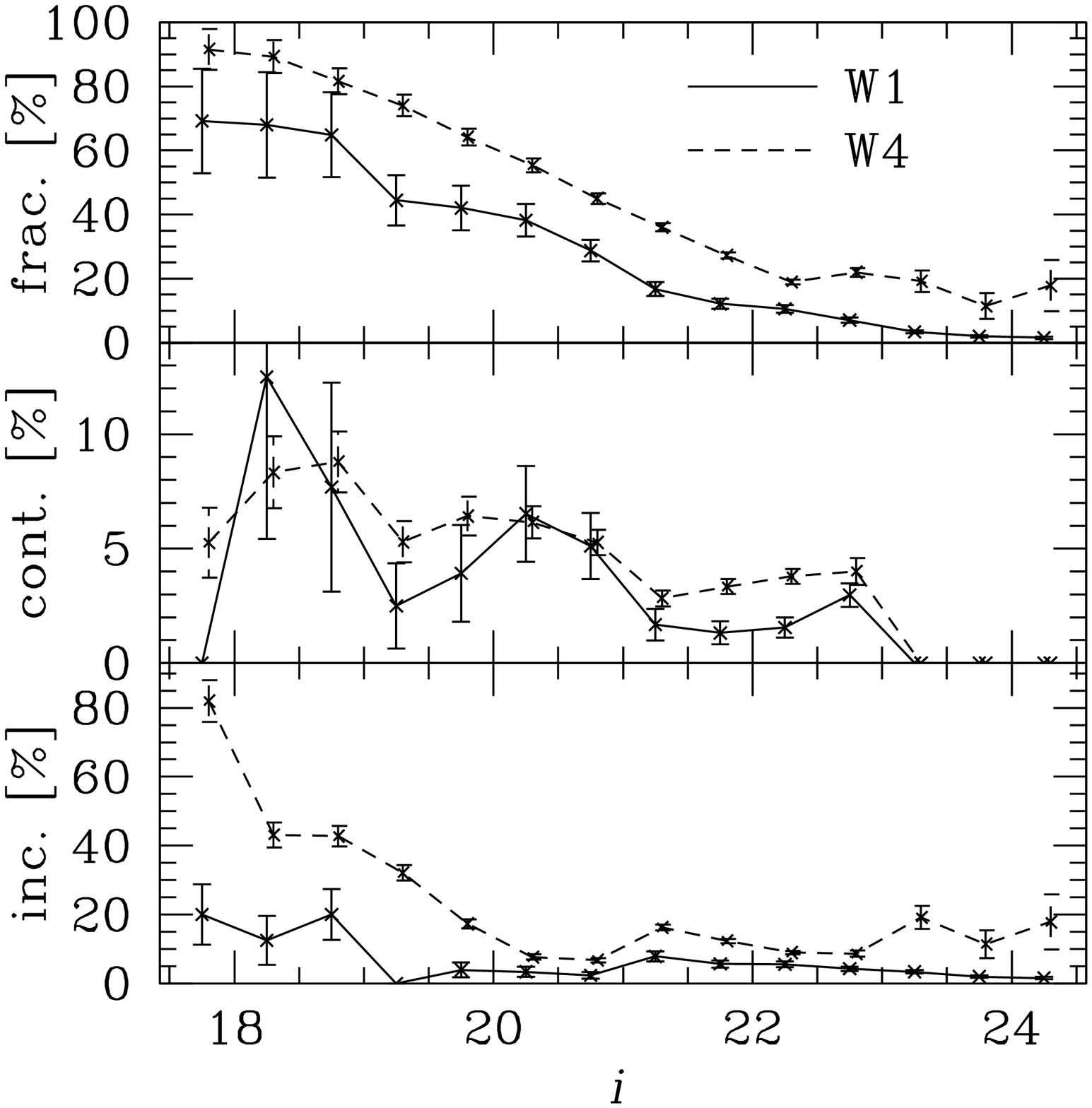}
\includegraphics[width=0.45\textwidth]{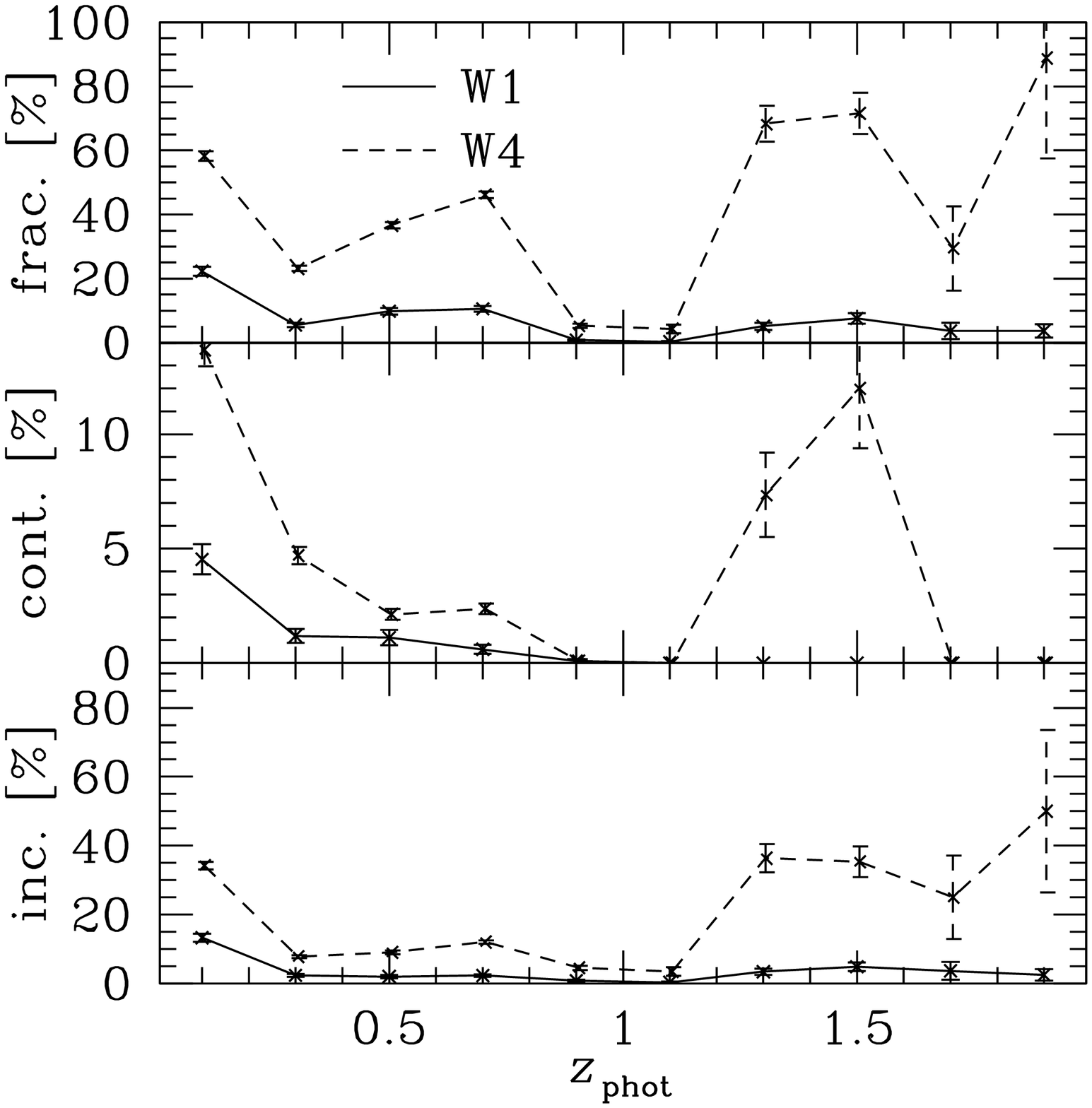}
\caption{Star-galaxy separation efficiency as a function of $i$-band
  magnitude (left) and photo-$z$ (right).  Shown are the fraction of
  stars (frac.; top), contamination of the galaxy sample (cont.;
  middle), and the incompleteness of the galaxy sample (inc.;
  bottom). Our estimator is tested and calibrated using the VVDS
  spectroscopic samples in W1 (solid line) and W4 (dashed
  line). Incompleteness represents the percentage of galaxies lost
  after selection, and contamination the percentage of stars
  misidentified as galaxies.}
\label{fig:star_galaxy}
\end{figure*}

We present the method efficiency as a function of magnitude and
redshift in Fig~\ref{fig:star_galaxy}. True galaxies and true stars
are given by spectroscopic information. The incompleteness is defined
as the percentage of galaxies lost after selection compared to the
total number of true galaxies, and the contamination as the number of
true stars misidentified as galaxies compared to the total number of
true galaxies.  A robust estimator should lead vanishing numbers for
both estimators.  In the range $21 < i < 23$, where size, magnitude
and colour information are used, the estimator performed the best,
keeping the contamination below 5 per cent and the incompleteness
below 15 per cent. Extending the estimator to brighter magnitude in
W4, where the star concentration is very high, increases the galaxy
incompleteness.  The star-galaxy separation strongly depends on the
redshift estimate of the objects. As seen in Sec.~\ref{sec:accuracy},
the redshift estimator is the most robust in the range $0.1 < z <
1.3$, where both incompleteness and contamination remain below 5 per
cent and 10 per cent, respectively. It seems however that a
conservative redshift cut $z < 1.1$ should be observed if a very
complete galaxy sample is needed in star-crowded fields, or
alternatively apply a hard size cut of $r_h>r_{h,{\rm limit}}$ so that
no star will pass. It should be noted that the high-galactic-latitude
fields of the CFHTLS contain only very few stars so that star-galaxy
separation is not a major issue for most WL science projects planned
by the CFHTLenS team.

\subsection{Redshift distributions}
\label{sec:z_dist}
\begin{figure*}
\includegraphics[width=0.32\textwidth]{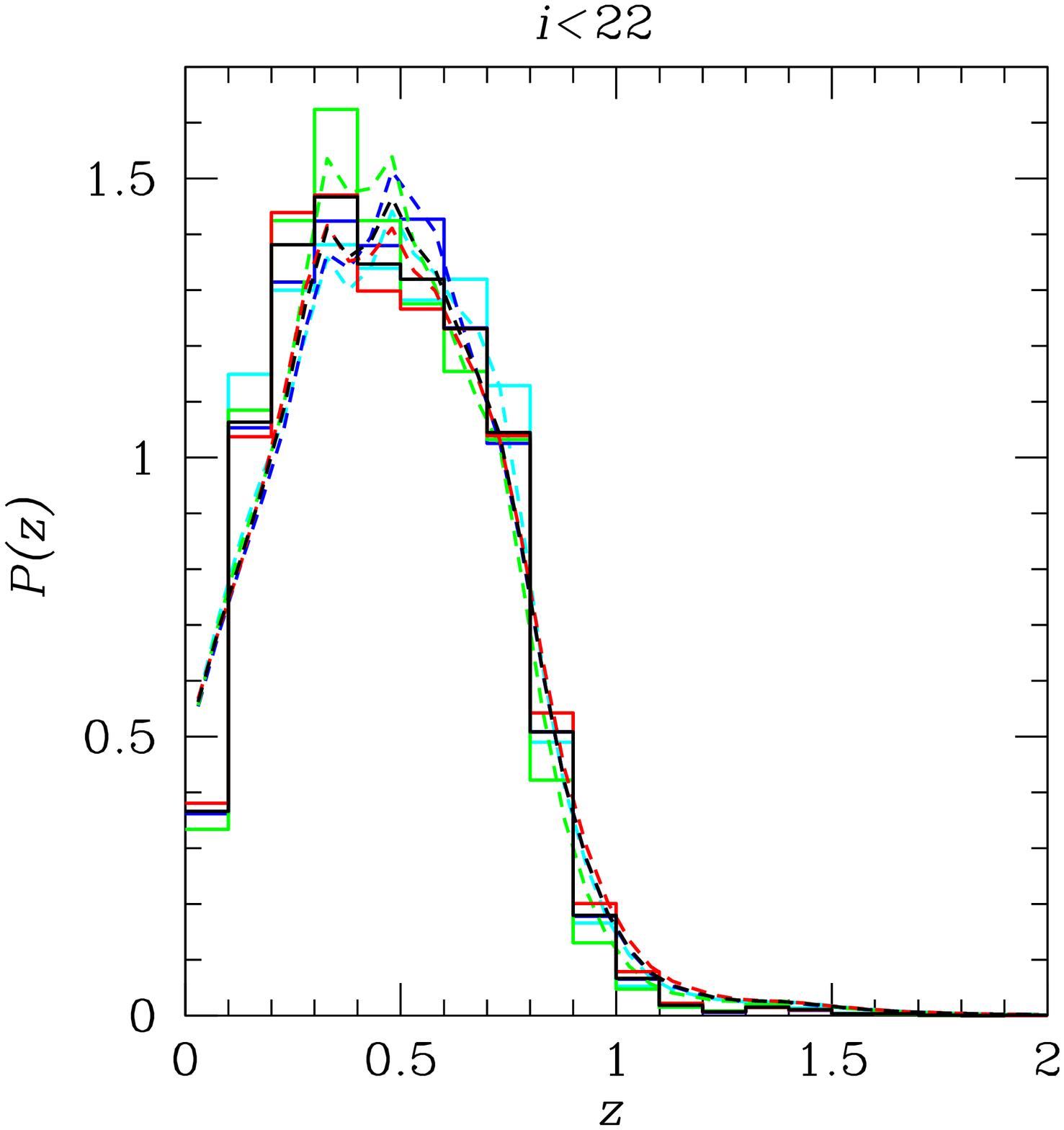}
\includegraphics[width=0.32\textwidth]{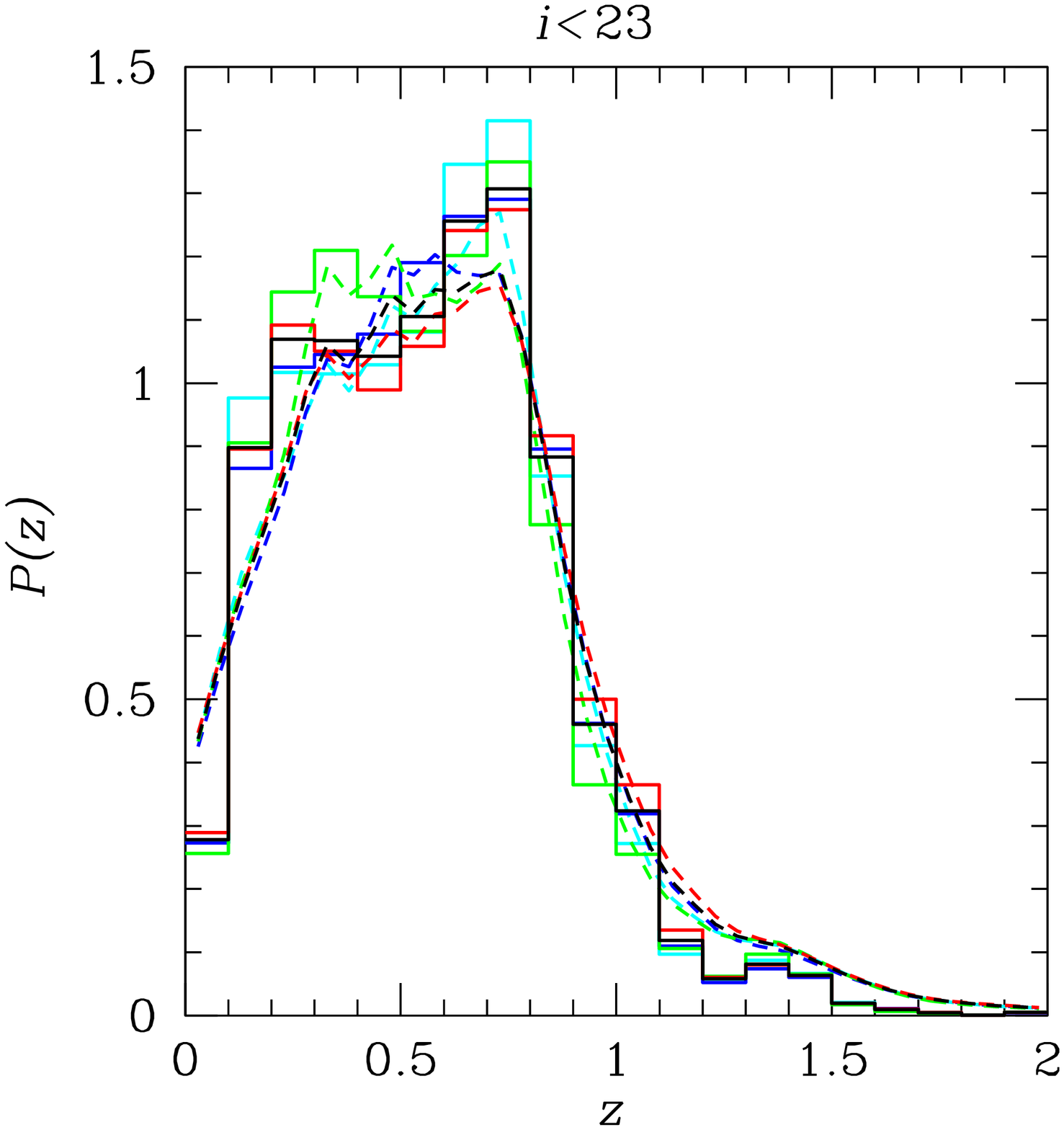}
\includegraphics[width=0.32\textwidth]{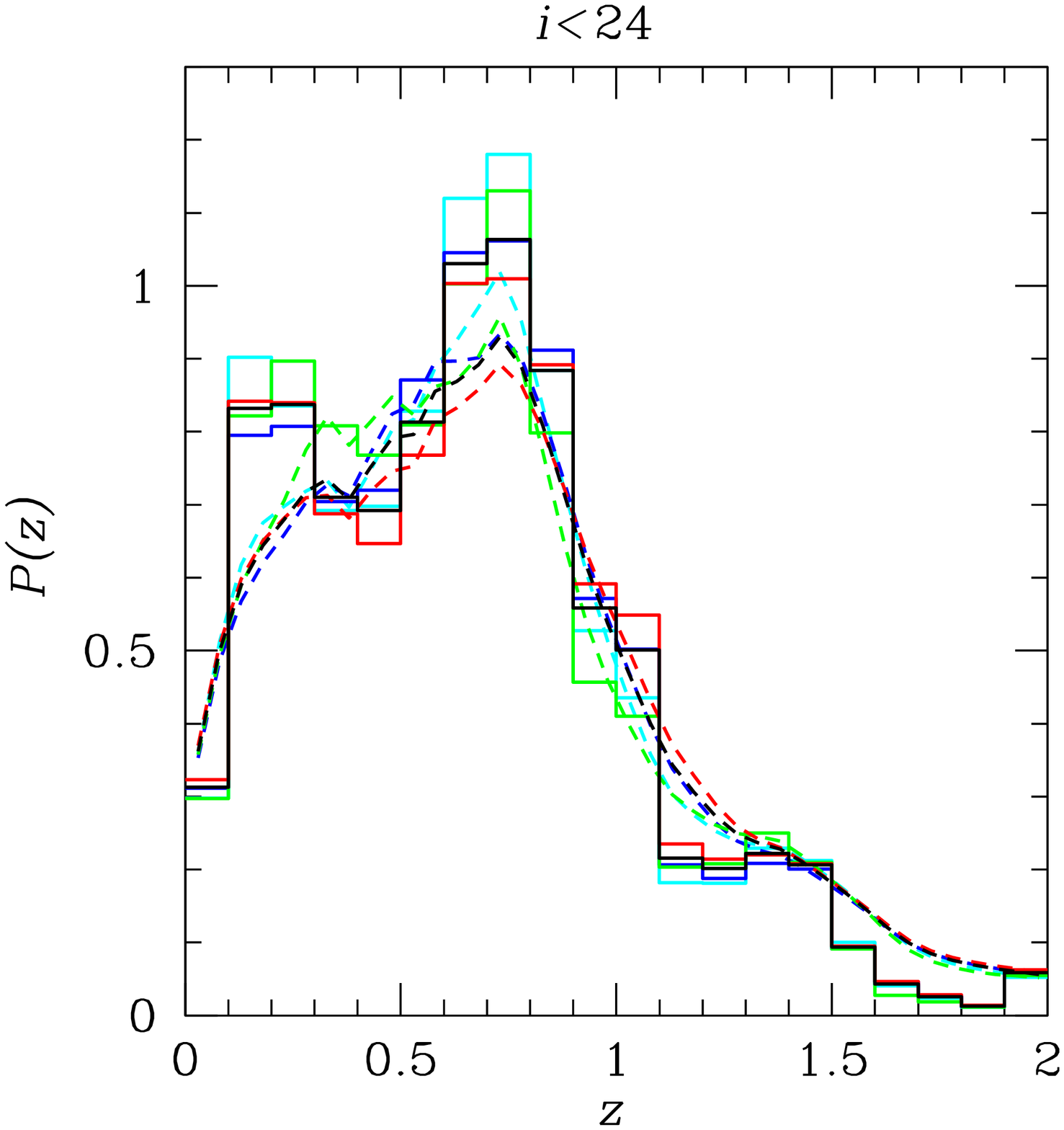}
\caption{Redshift distributions for $i<22$ (\emph{left}), $i<23$
  (\emph{middle}), and $i<24$ (\emph{right}). The dashed lines
  represent the stacked PDFs and the solid histograms represent the
  distributions of the most probable photo-$z$ values. The whole Wide
  survey is shown in black, W1 in red, W2 in green, W3 in blue, and W4
  in cyan.}
\label{fig:z_dist}
\end{figure*}

The redshift distributions of all objects classified as galaxies are
shown in Fig.~\ref{fig:z_dist} for three different magnitude
limits. We show the distributions of the most probable photo-$z$'s as
well as the stacked posterior probabilities output by the photo-$z$
code. While there is good agreement between the two for $i<23$, the
double-peaked histogram for the faintest magnitude cut ($i<24$) is not
reproduced in the stacked posterior probabilities. This
redshift-focussing effect occurs when the prior dominates the
posterior probability distribution for wide, flat likelihoods
(plateaus) in the low S/N case. A large number of objects are then
assigned the peak value of the prior which leads to artificial peaks
in the redshift histograms. This is the reason why we recommend to use
the full probability distributions, after a proper deconvolution
taking the photometric uncertainties into account was performed,
instead of the most probable redshifts (i.e. the peak of the
posterior) in science analyses.\footnote{It should be noted that the
  stacked posterior probabilities are certainly affected by the ad-hoc
  modification of the prior described in Sect.~\ref{sec:prior}.}
Similar double-peaked photo-$z$ distributions for $i<24$ in the CFHTLS
can be seen in \cite{2006A&A...457..841I} and
\cite{2009A&A...500..981C} so we suspect the filter set plays a
crucial role here. But also the prior can lead to multi-peaked
distributions. A final answer to this question requires a highly
complete spec-$z$ catalogue all the way down to $i=24$.

There are several methods discussed in the literature to correct these
redshift probability densities
\citep[e.g.][]{2008ApJ...684...88N,2010MNRAS.408.1168B} to make them
more realistic using angular cross-correlation functions between
different photo-$z$ bins or between photo-$z$ and spec-$z$ samples. We
defer such an analysis carried out with the CFHTLenS redshifts to a
forthcoming paper (Benjamin et al. in prep.).

We intentionally do not present functional fits to these
distributions.  Contemporary WL surveys (like the CFHTLS) have reached
such a precision that redshift distributions taken from external
surveys would dominate the total error budget on most cosmological WL
measurements \citep{2006APh....26...91V}. It is one great advantage of
the CFHTLS over previous surveys to have a photo-$z$ estimate for each
galaxy used in the WL analysis.

\section{Conclusions}
\label{sec:conclusions}
The CFHTLS represents the largest and therefore most powerful WL
survey to date. Here we present new, improved photometric catalogues
and photo-$z$ which will be used for the work of the CFHTLenS team. We
show that the correction for PSF-effects in data taken in different
bands at different times is crucially important. Not only does the
accuracy of the resulting photo-$z$'s increase significantly if
PSF-effects are thoroughly corrected for, but also the overall
photometric homogeneity of a survey can be improved considerably by
employing such corrections. We show that this can be done up to the
point, where re-calibrations of the photometric zeropoints with
sub-samples of galaxies with spec-$z$'s become largely unnecessary.

This implies that the re-calibration of photometric zeropoints
described in the literature are mostly corrections for PSF-effects. It
is hence dangerous to assume that the photo-$z$ accuracy measured on
the same spec-$z$ sample that was also used for the re-calibration
will be matched by the photometric sample. On the positive side, these
findings suggest that future surveys like KiDS, DES, LSST, or Euclid
will benefit tremendously from careful PSF homogenisation. These
projects will not necessarily need a spectroscopic coverage over the
whole area to achieve their absolute photometric calibration goals, if
PSF-effects are corrected for with high precision. For example, Euclid
will calibrate its redshift distributions directly from very complete
spectroscopic catalogues that will not cover the whole area. Relating
the results from these calibration fields to the rest of the survey
requires exquisite photometric homogeneity.

The CFHTLenS team will use these photo-$z$ catalogues for a wide
variety of WL-related science projects ranging from galaxy-galaxy
lensing and cluster lensing to cosmic shear tomography using the shear
as well as the magnification effect of WL. Especially the cosmological
measurements will benefit from this very homogeneous photo-$z$
catalogue. Measuring tiny correlations over large angular distances,
as it is done in cosmic shear or cosmic magnification studies,
requires an exquisite level of control of systematic effects. The
removal of PSF-effects from the photometry and the resulting accurate
photo-$z$'s represent an important step to reach this goal.

The shear measurement technique of CFHTLenS will be presented in
another technical paper (Miller et al. in prep.) and the resulting
shear catalogue will be carefully inspected for systematic effects
(Heymans et al. in prep.). Together with the photo-$z$'s presented in
this technical paper this will set the basis for CFHTLenS science
analyses.

\section*{Acknowledgements}

We are grateful to the CFHTLS survey team for conducting the
observations and the TERAPIX team for developing software used in this
study. We acknowledge use of the Canadian Astronomy Data Centre
operated by the Dominion Astrophysical Observatory for the National
Research Council of Canada's Herzberg Institute of Astrophysics.

We would like to thank Mischa Schirmer for his work on the THELI
pipeline as well as Stephane Arnouts, Jean-Charles Cuillandre, Yuliana
Goranova, Patrick Hudelot, Olivier Ilbert, and Henry J. McCracken for
their work on the TERAPIX T0006 catalogues.

The computational infrastructure for this project was supported by an
NSERC RTI grant as well as the Marie Curie IOF 252760.

H. Hildebrandt is supported by the Marie Curie IOF 252760 and by a
CITA National Fellowship. TE is supported by the BMBF through project
"GAVO III" and by the DFG through project ER 327/3-1 and the TR 33.
LVW is supported by NSERC and CIfAR. CH acknowledges support from the
European Research Council under the EC FP7 grant number 240185. JC is
supported by the Japanese Society for the Promotion of Science. CB is
supported by the Spanish Science MinistryAYA2009-13936,
Consolider-Ingenio CSD2007-00060, project2009SGR1398 from Generalitat
de Catalunya and by the the EuropeanCommission’s Marie Curie Initial
Training Network CosmoComp (PITN-GA-2009-238356). LF is supported by
the NSFC grants number 11103012 \& 10878003, Innovation Program
12ZZ134 and Chen Guang project 10CG46 of Shanghai Municipal Education
Commission, and Science and Technology Commission of Shanghai
Municipal Grant No. 11290706600. TDK was supported by an RAS 2010
Fellowship. MV acknowledges support from the Netherlands Organization
for Scientific Research (NWO) and from the Beecroft Institute for
Particle Astrophysics and Cosmology. H. Hoekstra and ES acknowledge
support from an NWO Vidi grant and a Marie Curie IRG. BR acknowledges
support from the European Research Council in the form of a Starting
Grant with number 240672.  TS acknowledges support from NSF through
grant AST-0444059-001, the Smithsonian Astrophysics Observatory
through grant GO0-11147A, and NWO.

\small

{\bf Author contributions:}

H. Hildebrandt led this paper and the photometry working group,
created the catalogues, estimated photo-$z$'s, and conducted the main
analysis.

TE reduced the complete data set and developed code for different
stages of the photometry and photo-$z$ estimation.

KK developed the shapelet-based code for the Gaussianisation of the
PSF.

LVW and CH managed and supervised the CFHTLenS collaboration and the
data flow. LVW also manually selected stars from size-magnitude
diagrams of $>6000$ chips of the MEGACAM mosaic.

JC led the photometric star-galaxy separation and contributed to the
T0006 photo-$z$'s.

JB, CB, LF, LM, and MV ran different tests for systematic effects in the
data (see Sect.~\ref{sec:quality_control}).

H. Hoekstra, TK, YM, MH, BR, TS, and ES contributed to this paper and
the photo-$z$ catalogue by organising the CFHTLS survey operations,
using the photo-$z$'s, providing valuable feedback, and engaging in
numerous discussion on how to improve the catalogue.

NB developed the \emph{BPZ} photo-$z$ code, which was used extensively in
this study.

\normalsize

\bibliographystyle{mn2e_mod}

\bibliography{photo_z_CFHTLenS}

\appendix

\section{Details of the PSF homogenisation}
\label{appA}
A key part of our analysis is the homogenisation of the PSF between
images taken through different filters, so that proper colours,
representing the same part of each source, can be measured. Our
approach is to convolve the images with a kernel that renders the PSF
close to Gaussian, with a width that is set by the worst-seeing image.

The PSF Gaussianisation is performed by first constructing a suitable,
spatially variable, kernel, and then convolving the images with
it. Both steps take advantages of some of the mathematical properties
of the shapelet formalism \citep{2003MNRAS.338...35R}.

Shapelets are two-dimensional Gauss-Hermite functions, of the form
\begin{equation}
S_{ab}(x,y)=N_{ab}H_a(x/\beta)H_b(y/\beta)e^{-r^2/2\beta^2} . \label{eq:shpl}
\end{equation}
$N_{ab}$ is a normalisation constant, and $\beta$ the scale
radius. $x$ and $y$ are Cartesian sky coordinates with respect to a
suitably chosen centre; $r$ is the corresponding polar
coordinate. $H_n$ is a Hermite polynomial.

Any source with intensity $I(x,y)$ can be written as a superposition
of shapelets:
\begin{equation}
I(x,y)=\sum_{ab}s_{ab}S_{ab}(x,y)
\end{equation}
where the $s_{ab}$ are the amplitudes of the different shapelets.

Shapelets have the useful property that a convolution of any two of
them can be written as a new (generally infinite) shapelet series:
\begin{equation}
S_{ab} \otimes S_{cd} = \sum_{ef} C_{ace}C_{bdf}S_{ef}
\end{equation}
and expressions for the matrix elements $C_{lmn}$ are given in
\cite{2003MNRAS.338...48R}. (Note that there is no requirement for the
scale radii of the shapelets to be the same.) If we express the PSF
and the kernel as shapelet series, with coefficients $p_{ab}$ and
$k_{ab}$ respectively, then this allows us to write the result of the
convolution $P\otimes K$ as a new shapelet series with coefficients
\begin{equation}
t_{ef}=\sum_{ab}\sum_{cd}C_{ace}C_{bdf}p_{ab}k_{cd}\equiv \sum_{cd}M_{cd,ef}k_{cd}
\label{eq:psfconv}
\end{equation}
with
\begin{equation}
M_{cd,ef}=\sum_{ab}C_{ace}C_{bdf}p_{ab}
\end{equation}
encoding the effect of PSF convolution in shapelet space. If the PSF
is known, $M$ can be computed.  Our Gaussianisation technique involves
constructing a shapelet kernel by inverting the
equation~\ref{eq:psfconv}. As target PSF on the left-hand side we
stipulate a Gaussian; all its elements are zero except the $t_{00}$
component.

Before calculating the kernel, we model the PSF variation across the
full CFHTLS image. This is done by identifying all stars above a
certain S/N ratio, choosing a suitable scale radius for the PSF,
making a flux-normalised shapelet expansion for each star, and then
separately fitting (with a two-dimensional polynomial) the variation
of each coefficient $s_{ab}$ across the image, with outlier rejection.

This PSF map is then sampled on a regular grid across the image, the
convolution kernel is calculated at each of those points, and a
polynomial model fitted to the variation of its coefficients across
the image. This kernel is then convolved with the original CFHT
image. Again here we can use a nice mathematical property of
shapelets: each is its own Fourier transform, making the convolution
efficient.

In practice we truncate the shapelet series: we only have pixellated
information on the PSF and so cannot sample it arbitrarily finely. For
robustness we truncate the kernel at a lower order than the PSF, to
prevent overfitting. The inversion of eq.~\ref{eq:psfconv} is
therefore recast as a least-squares problem, in which we determine the
$k_{cd}$ that best approximate the target Gaussian PSF $t_{ab}$.

For the CFHT data we find that the following parameters work well:
Shapelet order (maximum $a+b$ in the expansion eq.~\ref{eq:shpl})
equal to 10 for the PSF, 8 for the kernel; polynomial order for
fitting spatial variation across the image 5; shapelet scale radius of
the PSF map 1.3 times the median Gaussian radius fitted to the stars;
and target Gaussian radius 0.8 times the largest scale radius of the
images to be compared.

\label{lastpage}
\end{document}